\documentclass{aa}

\usepackage{graphicx}
\usepackage{txfonts}

\begin{document} 
	\titlerunning{Photo-z for X-ray AGN using machine learning techniques}
\authorrunning{Mountrichas et al.}

   \title{Estimating Photometric Redshifts for X-ray sources in the X-ATLAS field, using machine-learning techniques}

   \author{G. Mountrichas\inst{1}, A. Corral\inst{2,1}, V. A. Masoura\inst{1,3},  I. Georgantopoulos\inst{1}, A. Ruiz\inst{1}, A. Georgakakis\inst{1}, F. J. Carrera\inst{2}, S. Fotopoulou\inst{4}}
          
    \institute{National Observatory of Athens, V.  Paulou  \& I.  Metaxa, 11532,  Greece
              \email{gmountrichas@gmail.com}
         \and
             Instituto de Fisica de Cantabria (CSIC-Universidad de Cantabria), 39005 Santander, Spain
          \and
             Section of Astrophysics, Astronomy and Mechanics, Department of Physics, Aristotle University of Thessaloniki, 54 124, Thessaloniki, Greece 
          \and
             Department of Astronomy, University of Geneva, ch. d'Ecogia 16, 1290, Versoix, Switzerland}

 \abstract
 {We present photometric redshifts for 1,031 X-ray sources in the X-ATLAS field, using the machine learning technique TPZ (Kind2013). X-ATLAS covers 7.1\,deg$^2$ observed with the XMM-{\it{Newton}} within the Science Demonstration Phase (SDP) of the H-ATLAS field, making it one of the largest contiguous areas of the sky with both XMM{\it{Newton}} and {\it{Herschel}} coverage. All of the  sources have available SDSS photometry while 810 have additionally mid-IR and/or near-IR photometry. A spectroscopic sample of 5,157 sources primarily in the XMM/XXL field, but also from several X-ray surveys and the SDSS DR13 redshift catalogue, is used for the training of the algorithm. Our analysis reveals that the algorithm performs best when the sources are split, based on their optical morphology, into point-like and extended sources. Optical photometry alone is not enough for the estimation of accurate photometric redshifts, but the results greatly improve when, at least, mid-IR photometry is added in the training process. In particular, our measurements show that the estimated photometric redshifts for the X-ray sources of the training sample, have a normalized absolute median deviation, nmad$\approx$0.06, and the percentage of outliers, $\eta$=10-14\%, depending on whether the sources are extended or point-like. Our final catalogue contains photometric redshifts for 933 out of the 1,031 X-ray sources with a median redshift of 0.9. }


   \maketitle

%
%-------------------------------------------------------------------

\section{Introduction}

   Current and future surveys (e.g. XMM, eROSITA, DES, Euclid) will provide us with large datasets  that contain hundreds of thousands of sources. Spectroscopy is expensive in telescope time and challenging to complete for large samples, thus photometric redshift (photo-z) estimations have become a necessity in observational astronomy nowadays. Although photo-z estimations are cheaper and the only mean to estimate distances for large samples, they are also subject to systematics and higher uncertainties compared to spectroscopic redshift estimations (spec-z).

The pursuit of accurate photometric redshifts has led to the development of many photo-z estimation methods that can be divided into two main categories: template fitting  \citep[e.g.][]{Brammer2008} and machine learning \citep[e.g.][]{Kind2013} techniques, although there are some hybrid \citep[e.g.][]{Beck2017} ones, as well. The template fitting techniques determine the photometric redshifts by fitting synthetic spectral templates, either empirical ones or synthesized from stellar population models to observational spectral templates. A number of variations of this technique exist in the literature, e.g. Bayesian Photometric Redshifts \citep[BPZ;][]{Benitez2000}, Easy and Accurate Redshifts from Yale \citep[EAZY;][]{Brammer2008}. Machine-learning techniques, also known as empirical methods, use a spectroscopic dataset to train an algorithm and then the trained algorithm is applied to a photometric sample to estimate photometric redshifts. Examples of empirical methods include the Artificial neural network \citep[ANNz;][]{Collister2004, Lahav2012} and random forest techniques, e.g. Trees for Photo-Z \citep[TPZ;][]{Kind2013}.

Each of the aforementioned techniques has its own pros and cons. \cite{Beck2017} compared the performance of eight photo-z estimation methods (four template fitting techniques and four machine-learning techniques). Their analysis revealed that all methods perform adequately when the training set coverage is sufficient but their performance falls when extrapolation is required. Especially random forest techniques are not expected to perform well beyond the boundaries of the training set. On the other hand, the latter techniques perform better compared to the other techniques when the photometric measurement errors increase. \cite{Beck2017} concluded that none of the methods is superior and a trade-off has to be made depending on the available training set, i.e. its photometric accuracy and coverage.

The machine learning methods have been successfully applied for the derivation of photometric redshifts for galaxies 
\citep[e.g. SDSS;][]{Beck2016} and optical QSOs \citep[e.g.][]{Brescia2015, Cavuoti2017}. However, for X-ray AGN only SED fitting techniques have been used \citep[][]{Salvato2009, Hsu2014}. AGN SEDs though are more complicated than galaxies' SEDs due to e.g. contamination from the host galaxy, intrinsic obscuration, variability and dominance of different components in different spectral bands. Thus, photo-z for AGN via SED fitting is difficult. On the other hand, machine learning methods require large spectroscopic, training samples to perform well and X-ray datasets suitable to be used as training sets are sparse. 

In this work we use X-ray sources detected in the XMM-XXL survey \citep[][]{Liu2016} to train, for the first time, a machine learning algorithm \citep[TPZ;][]{Kind2013} to estimate photometric redshifts for X-ray AGN in the X-ATLAS field. Our goal is to use these photoz estimations in a future paper, to estimate the Star Formation Rate (SFR) and stellar mass of these sources and study the connection between the AGN activity and the environment of their host galaxy. In this paper, we will check the accuracy of the photo-z estimations. The structure of the paper is as follows: In Section 2 we describe the X-ray sources for which we estimate photo-z, in Section 3 we describe briefly the TPZ algorithm and provide information for the training sample. The results are presented in Section 4 while in Section 5 we discuss the and summarize the main conclusions of this work.

%--------------------------------------------------------------------
\section{The X-ray Sample}

The {\it{Herschel}} Terahertz Large Area survey (H-ATLAS) is the largest Open Time Key Project carried out with the {\it{Herschel}} Space Observatory \citep[][]{Eales2010}, covering an area of 550\,deg$^2$ in five far-infrared and sub-mm bands (100, 160, 250, 350 and 500\,$\mu$m). 16\,deg$^2$ have been presented in the Science Demonstration Phase (SDP) catalogue \citep[][]{Rigby2011} and lie within one of the regions observed by the Galaxy And Mass Assembly (GAMA) survey \citep[][]{Driver2011, Baldry2010}. XMM-{\it{Newton}} observed 7.1\,deg$^2$ with a total exposure time of 336ks (in the MOS1 camera) within the H-ATLAS SDP area, making the XMM-ATLAS one of the largest contiguous areas of the sky with both XMM-{\it{Newton}} and {\it{Herschel}} coverage. The catalogue contains 1816 unique sources \citep[][]{Ranalli2015}.

To obtain optical, mid-IR and far-infrared photometry for the XMM-ATLAS sources we cross-match the X-ray catalogue with the SDSS-DR13 \citep{Albareti2015}, the WISE \citep{wright2010} and the VISTA-VIKING catalogues \citep[][]{Emerson2006, Dalton2006}. The source matching was performed using the ARCHES cross-correlation tool xmatch, which matches symmetrically an arbitrary number of catalogues providing a Bayesian probability of association or non-association \citep{Pineau2016}. xmatch associates to each X-ray source one or more tuples including possible counterparts in VISTA and/or WISE, with the corresponding probability. If a given X-ray source has more than one associate tuple, we select those with probability >0.68 and, among those, those included in the most catalogues, and finally, those with the highest probability. The cross-match revealed 1031 sources with at least optical photometry. Using the association probabilities derived by xmatch, less than $10\%$ of the counterparts in our catalogue are missmatches ($\approx 85$ sources). 848 out of the 1031 sources have mid-infrared counterparts while 589 have also near-infrared (NIR) counterparts (Table \ref{table_agn_numbers}). Out of the 1031 sources, 174 have spectroscopic redshifts from either the SDSS or the GAMA surveys. 

%The colour distribution of the XMM-ATLAS sources is shown in Fig. \ref{fig_colours}.

\begin{table}
\caption{The number of X-ATLAS X-ray AGN divided based on their available photometry and optical morphology. In the parentheses we quote the number of sources with available spectroscopic redshift from the SDSS and GAMA surveys.}
\centering
\setlength{\tabcolsep}{0.7mm}
\begin{tabular}{cccc}
       \hline
 \hline
Available photometry & Total number of & Point-like & Extended \\
& sources &sources&sources \\
       \hline
SDSS & 1031 (174) & 576(119) & 455 (55) \\
SDSS+WISE  & 603 (124)  & 343 (87) & 260 (37)  \\
SDSS+WISE+NIR & 423 (92) & 249 (67) & 174 (25)  \\
SDSS+NIR& 653 (122) & 380 (86) & 273 (36)  \\
\hline
\label{table_agn_numbers}
\end{tabular}
\end{table}

\begin{figure}
\centering
\includegraphics[width=\hsize]{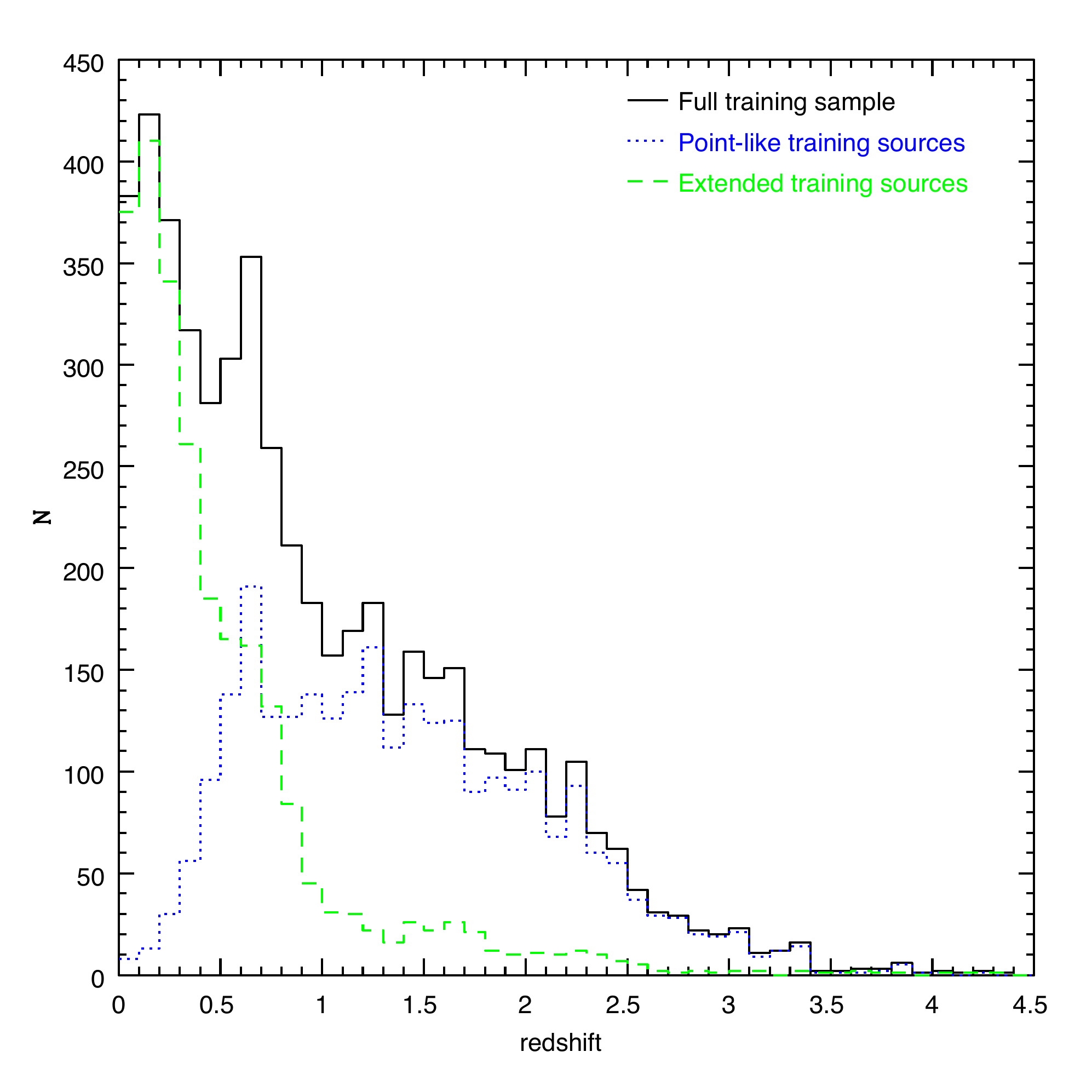}
\caption{The redshift distribution of the 5,157 sources used to train the TPZ algorithm (black, solid line). The dashed and dotted lines, present the redshift distribution when we split the training sources into extended and point-like, based on their optical classification.}
\label{fig_n_z}
\end{figure}

\begin{figure*}
\begin{center}
\includegraphics[height=0.7\columnwidth]{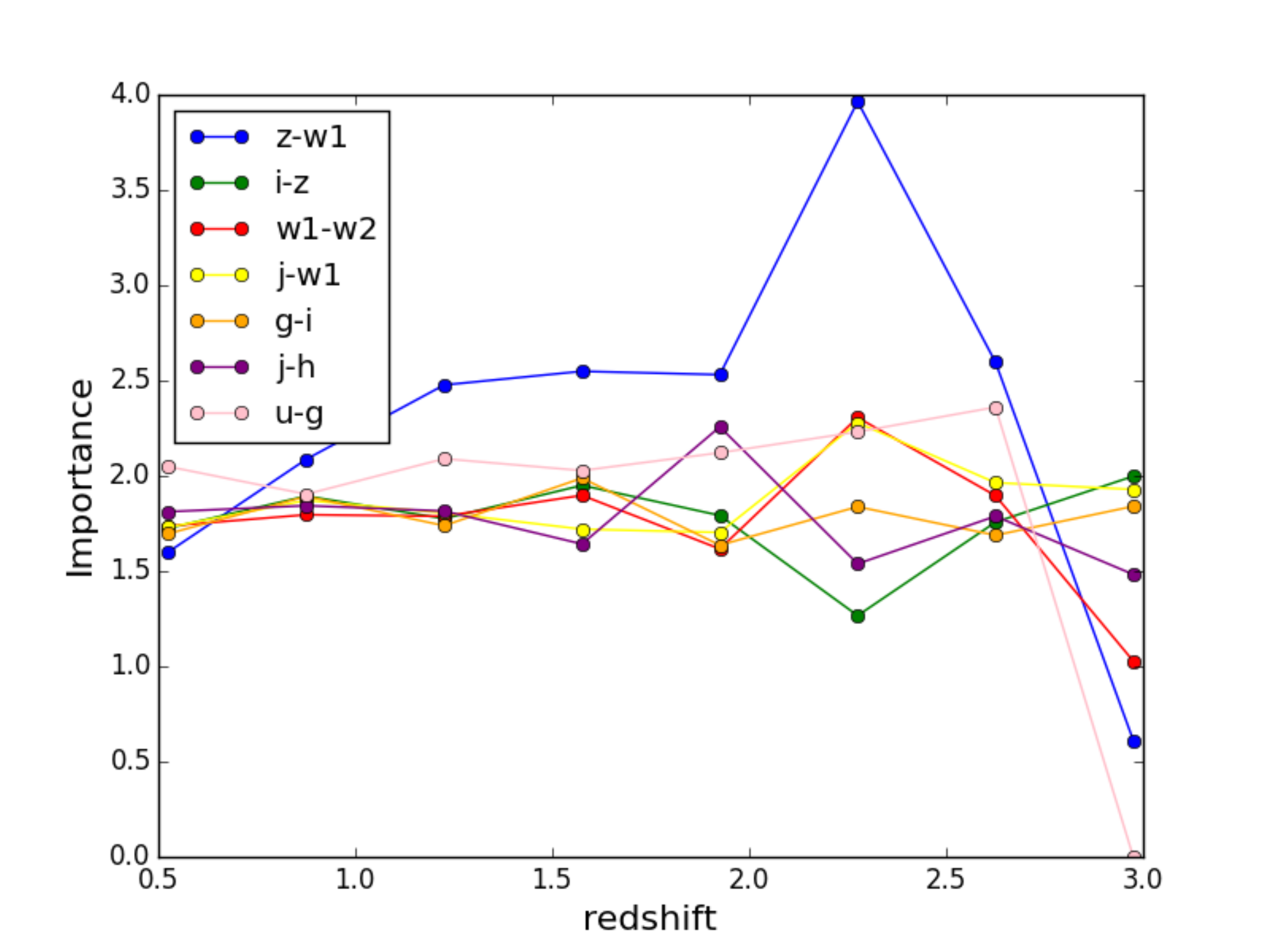}
\includegraphics[height=0.7\columnwidth]{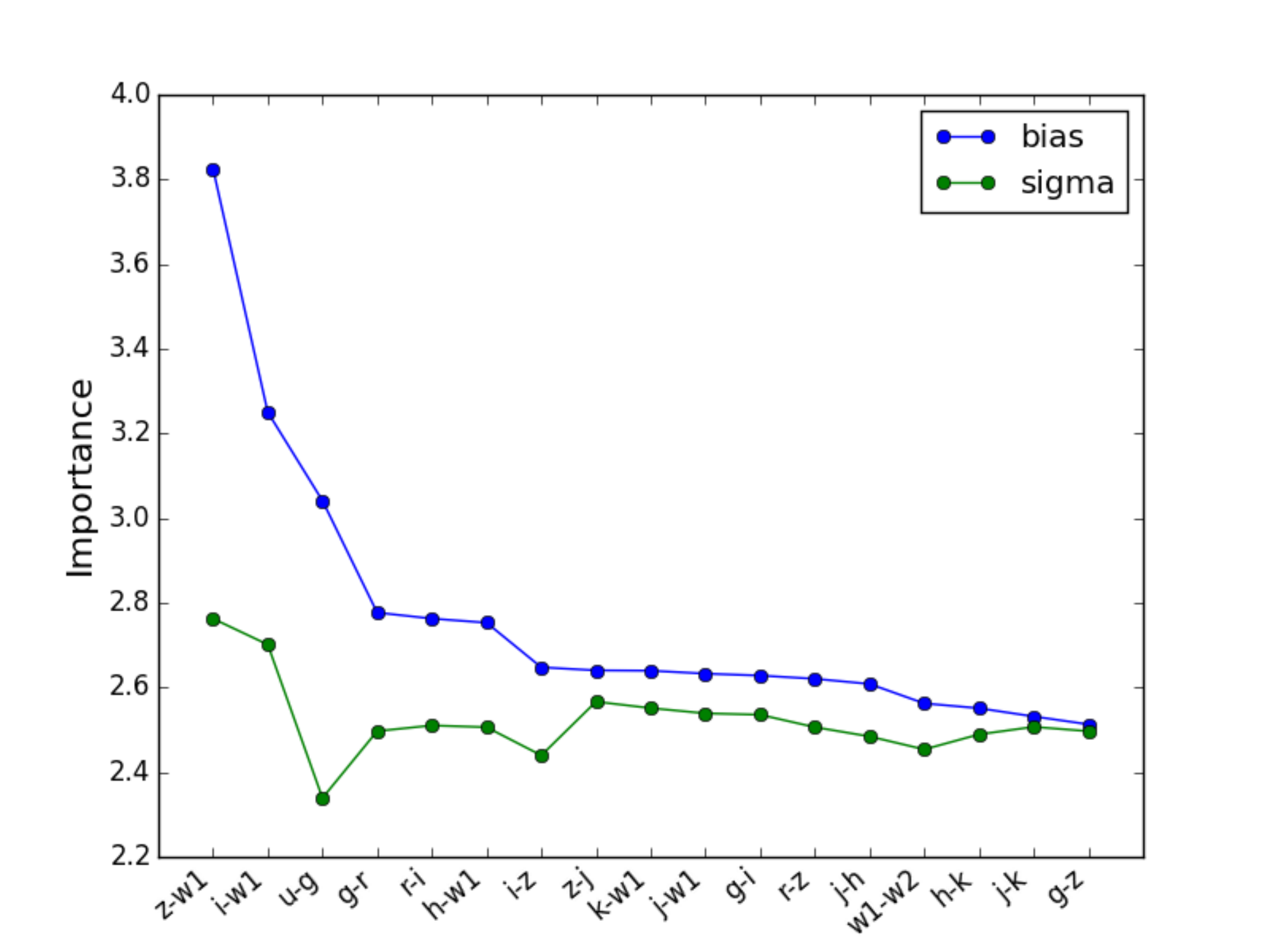}
\end{center}
\caption{Pointlike sources. {\bf{Left:}} Attributes importance as a function of redshift. {\bf{Right:}} The RMS importance factor as a function of the attributes computed by using the bias and its scatter.}
\label{fig_importance_point}
\end{figure*}

\begin{figure*}
\begin{center}
\includegraphics[height=0.7\columnwidth]{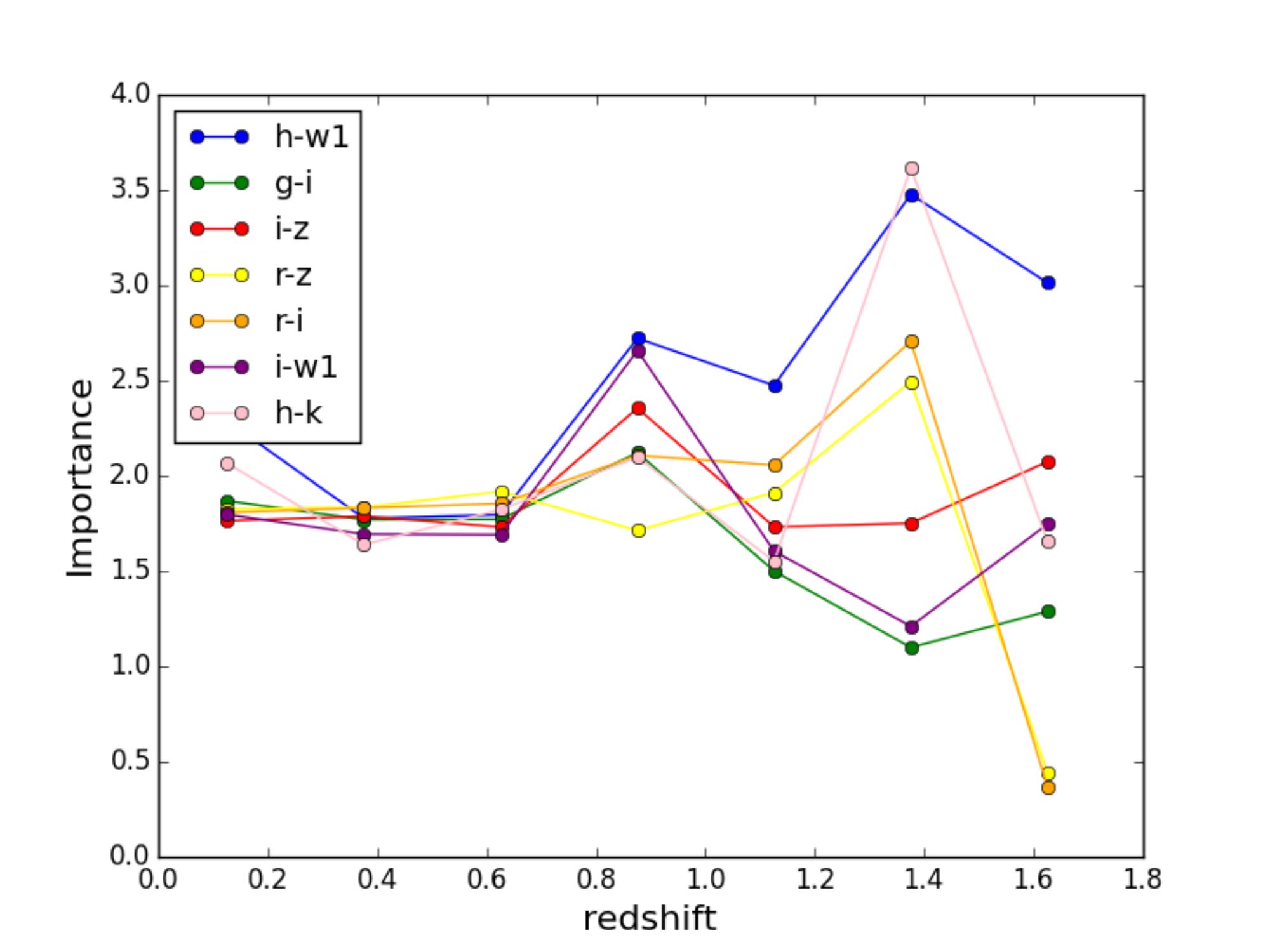}
\includegraphics[height=0.7\columnwidth]{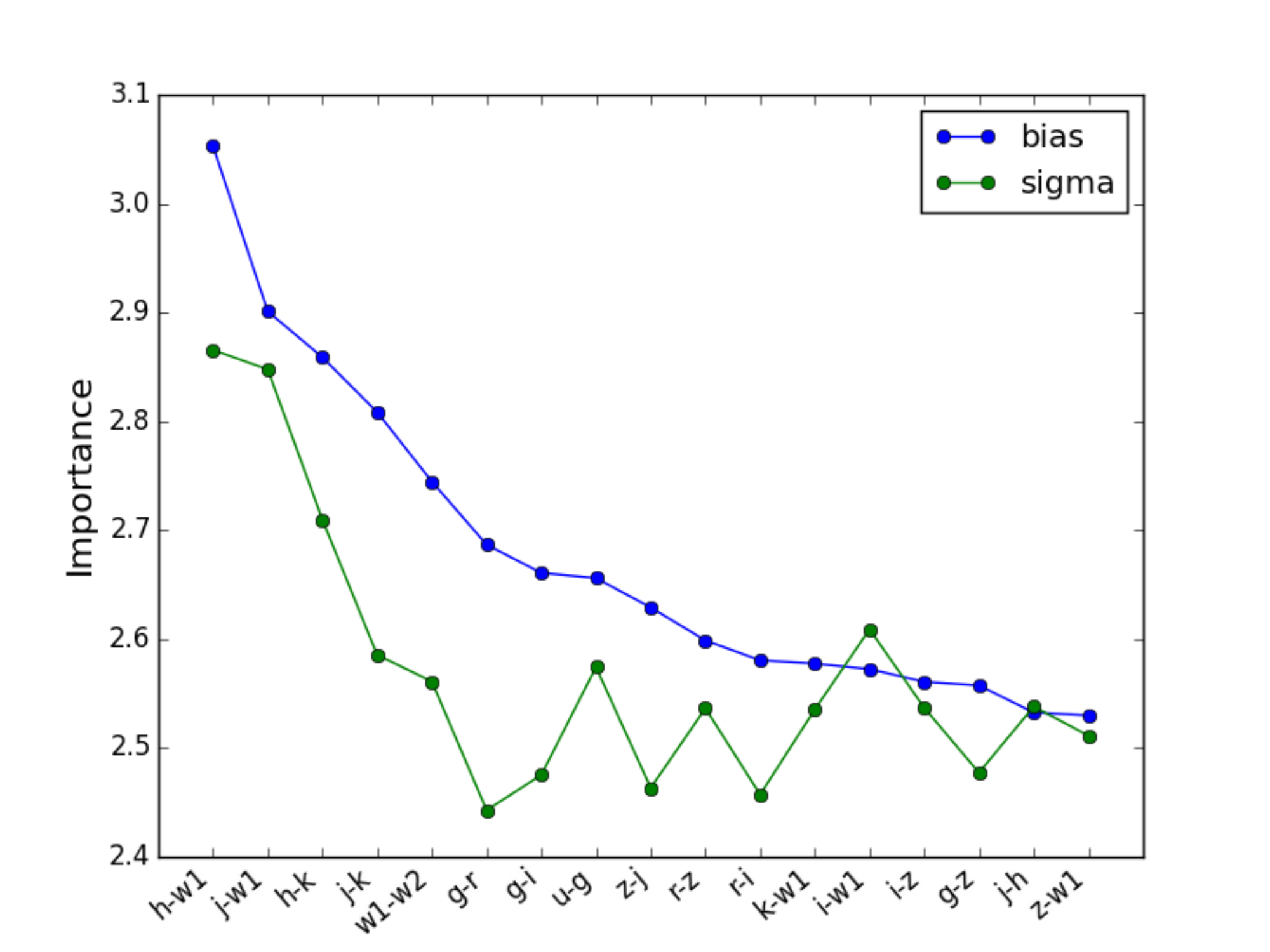}
\end{center}
\caption{Same measurements presented in Fig. \ref{fig_importance_point} but for extended sources.}
\label{fig_importance_extended}
\end{figure*}

\begin{figure*}
\centering
\includegraphics[height=1.\columnwidth]{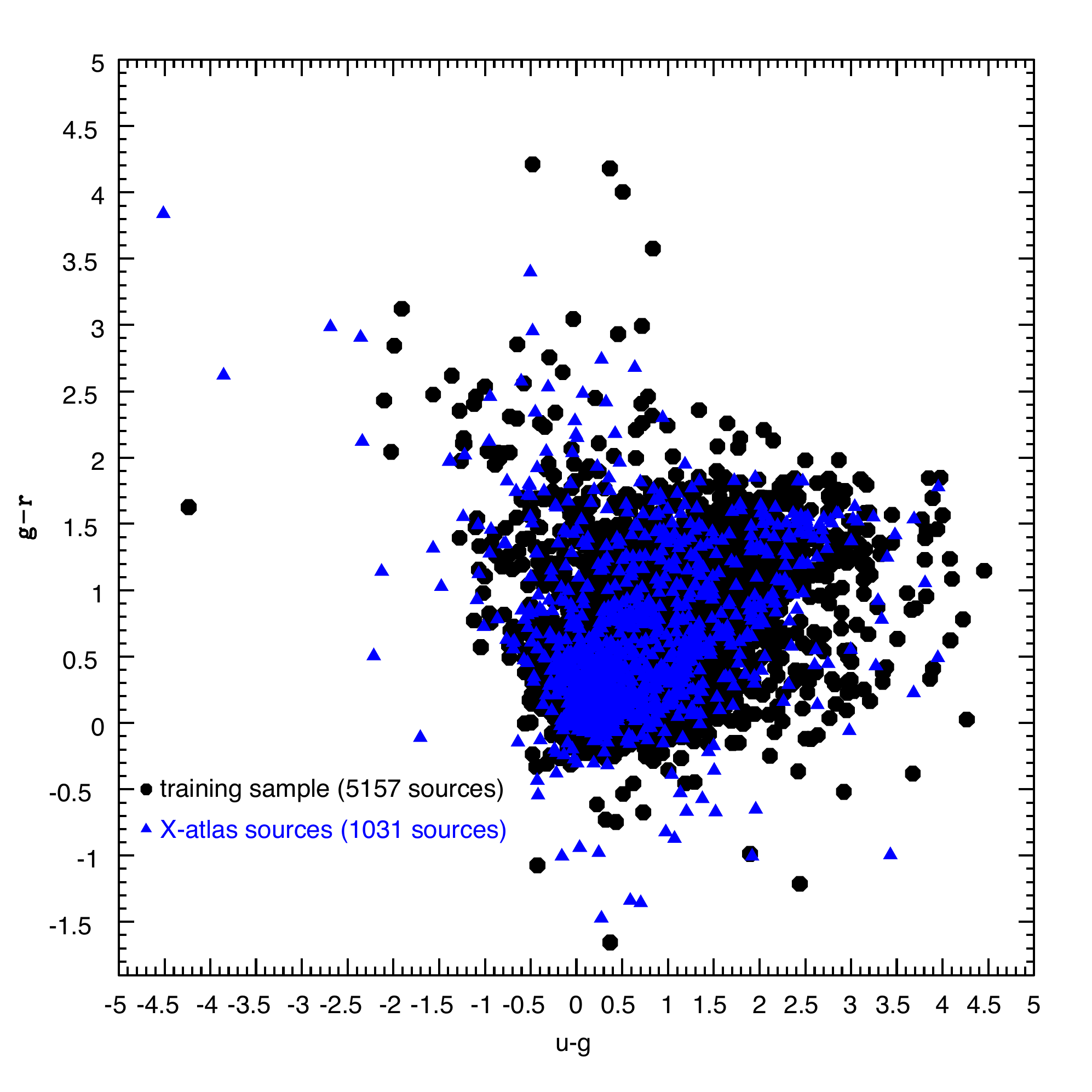}
\includegraphics[height=1.\columnwidth]{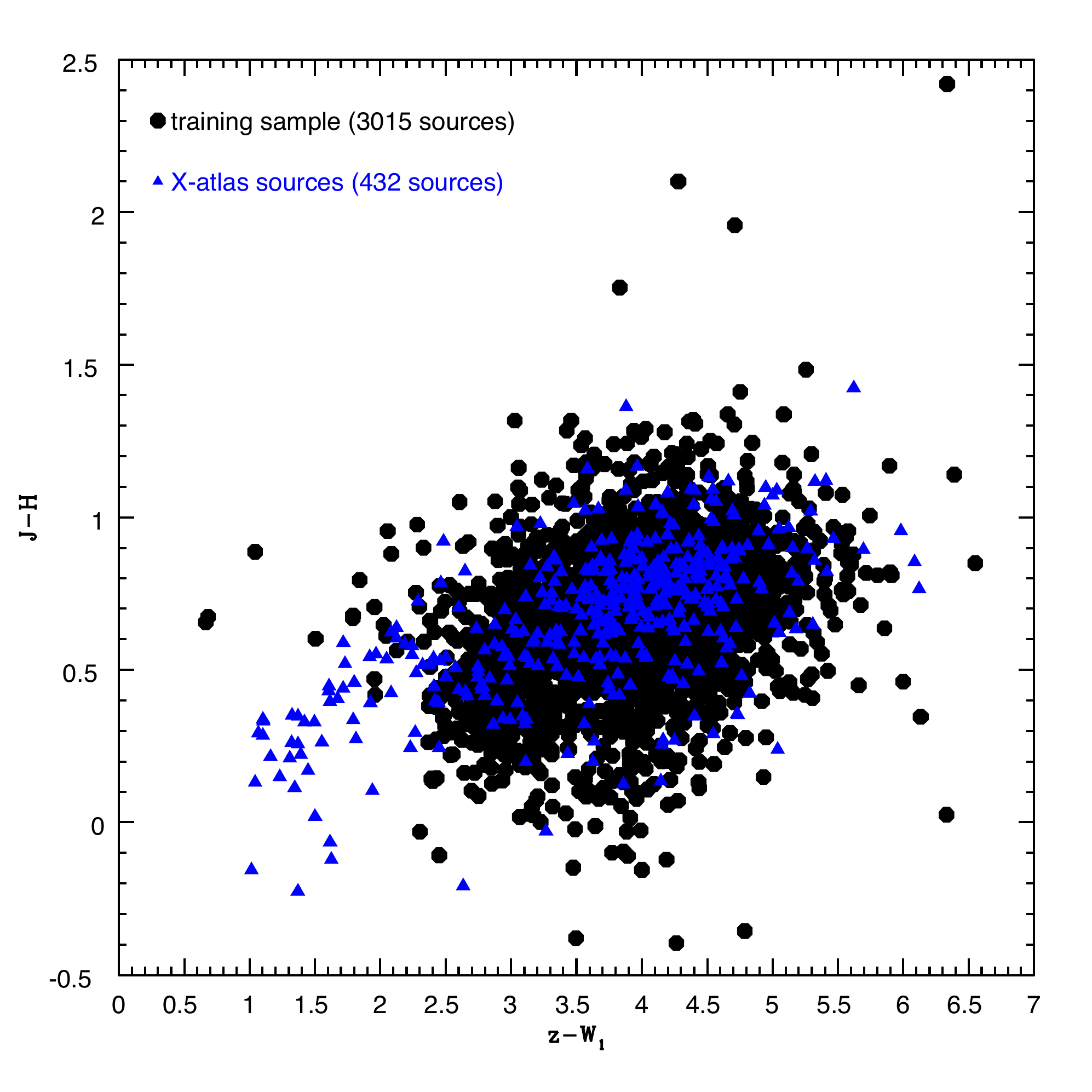}
\caption{{\bf  Left:} The u-g vs. g-r colour distribution of the training sample (black circles) and the X-ATLAS sources (blue triangles). {\bf Right:} The z-W1 vs. J-H colour distribution of the training sample (black circles) and the X-ATLAS sources (blue triangles). The fraction of the X-ATLAS sources that is well covered, by the training set, is different for different colour combinations. This is quantified in Table \ref{table_colours_kde}.}
\label{fig_colours}
\end{figure*}

\begin{figure*}
\begin{center}
\includegraphics[height=1.\columnwidth]{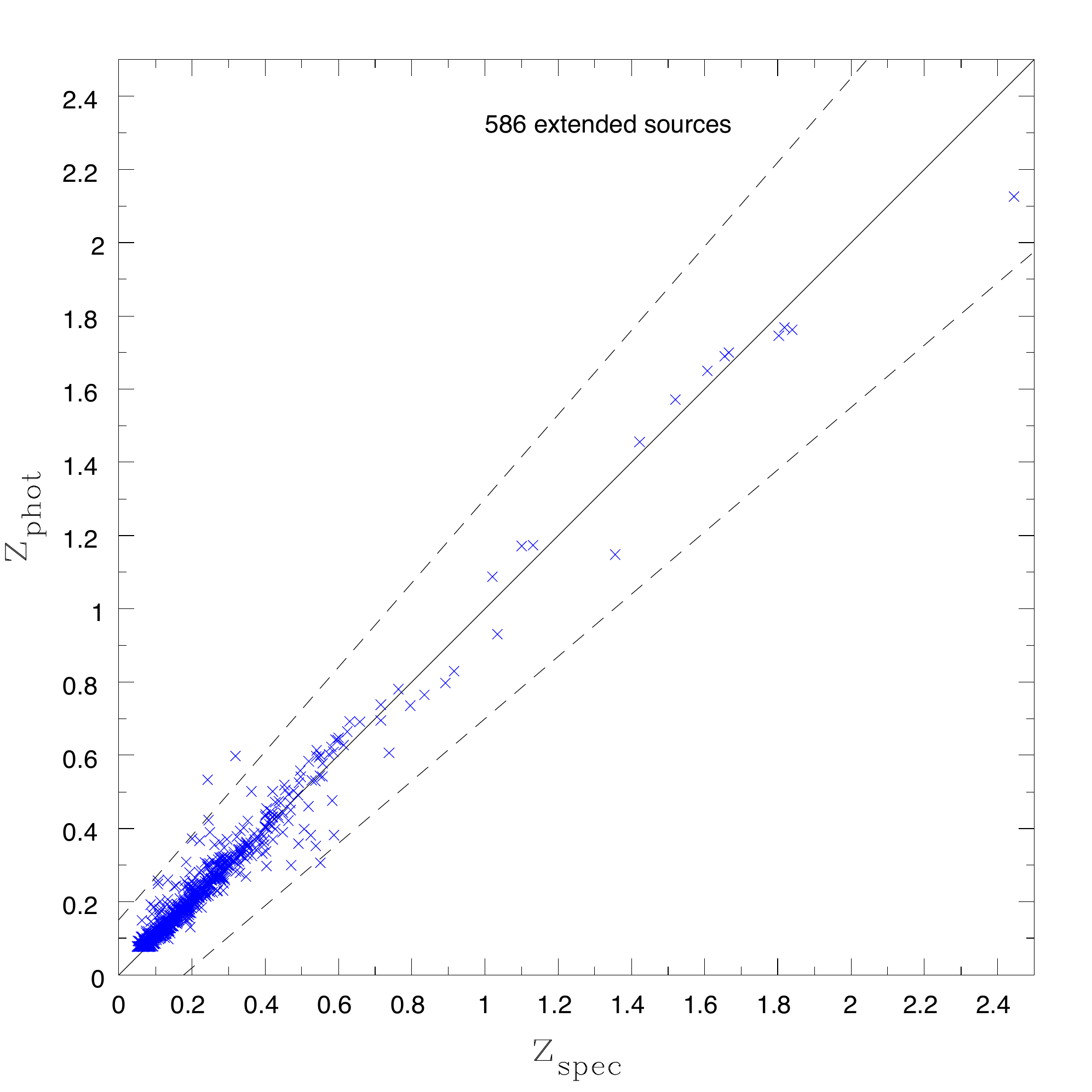}
\includegraphics[height=1.\columnwidth]{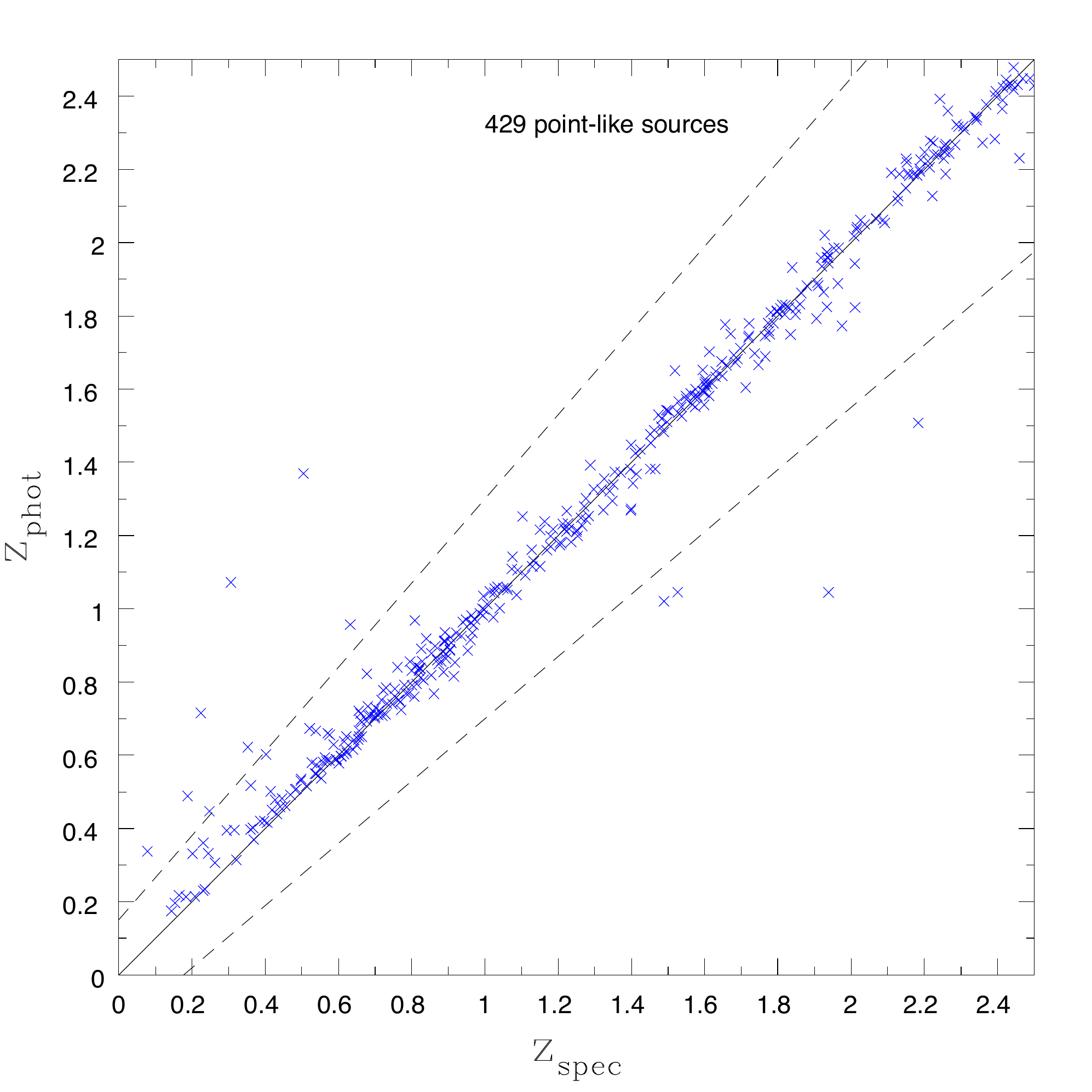}
\end{center}
\caption{The performance of TPZ using the ten available photometric bands (SDSS+WISE+near IR). The training sample has been split into train and test files to compare the estimated photometric redshifts with the spectroscopic redshifts of the sources. The dashed lines correspond to $\Delta \rm{z_{norm}}=\pm0.15$. Based on our analysis, the number of outliers is $\eta=9\%$ and $\eta=13\%$, for the extended and pointlike sources, respectively. The normalized absolute median deviation is $\sigma_{nmad}\approx 0.04-0.05$.}
\label{fig_test_tpz}
\end{figure*}

\begin{figure*}
\begin{center}
\includegraphics[height=0.95\columnwidth]{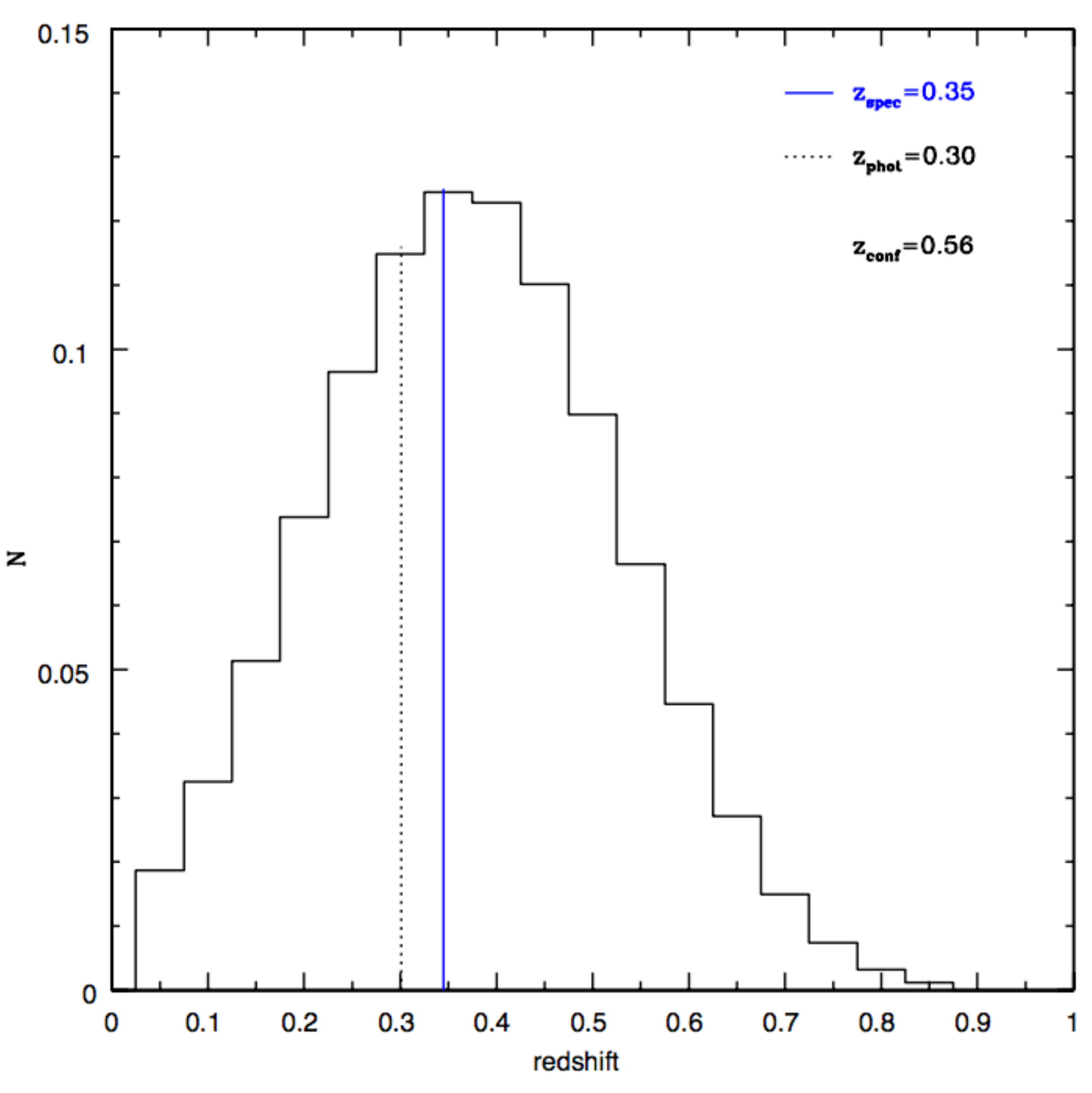}
\includegraphics[height=0.95\columnwidth]{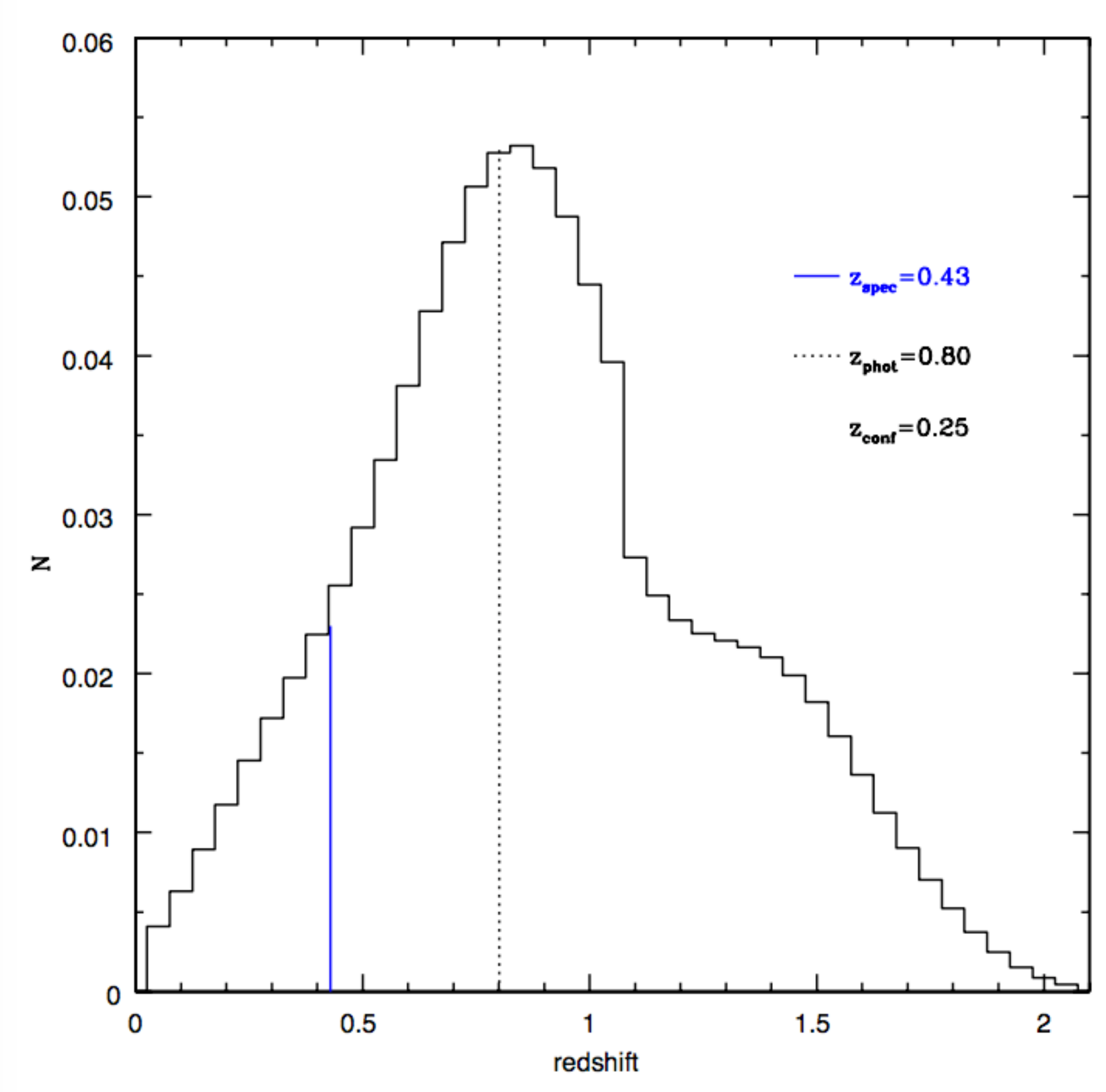}
\includegraphics[height=0.95\columnwidth]{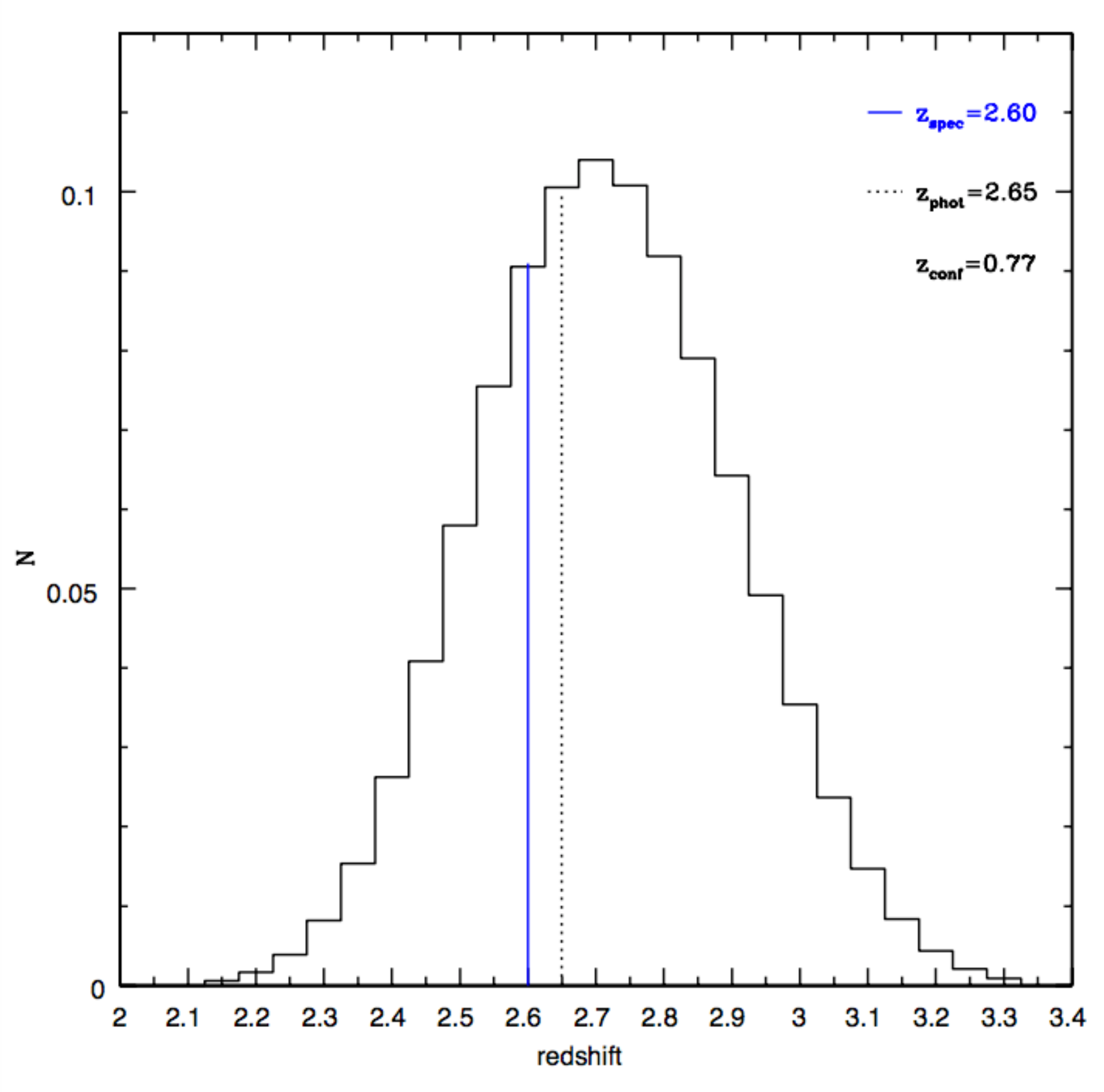}
\includegraphics[height=0.95\columnwidth]{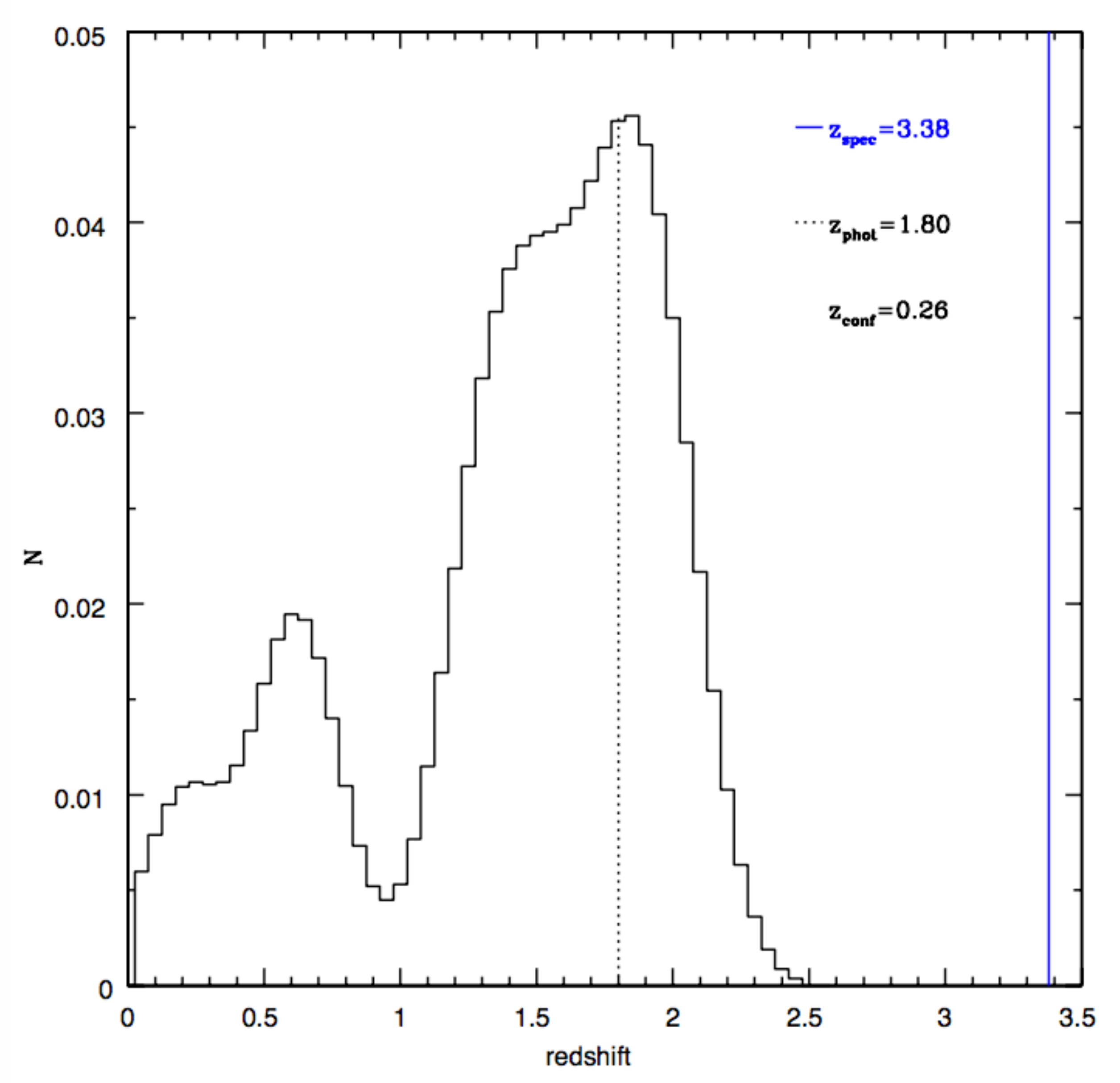}
\end{center}
\caption{Examples of PDFs produced by TPZ, during the validation process. The top panels present results for extended sources and the bottom panels for pointlike sources. On the left panels, the estimated photoz (dotted line) is in agreement with the spectroscopic redshift (solid line) of the source. On the right panels, the estimated photoz differs significantly compared to the spectroscopic redshift. These measurements are also characterized by low confidence level of the photometric redshift.}
\label{fig_pdf_examples}
\end{figure*}

\begin{figure*}
\centering
\includegraphics[height=1.\columnwidth]{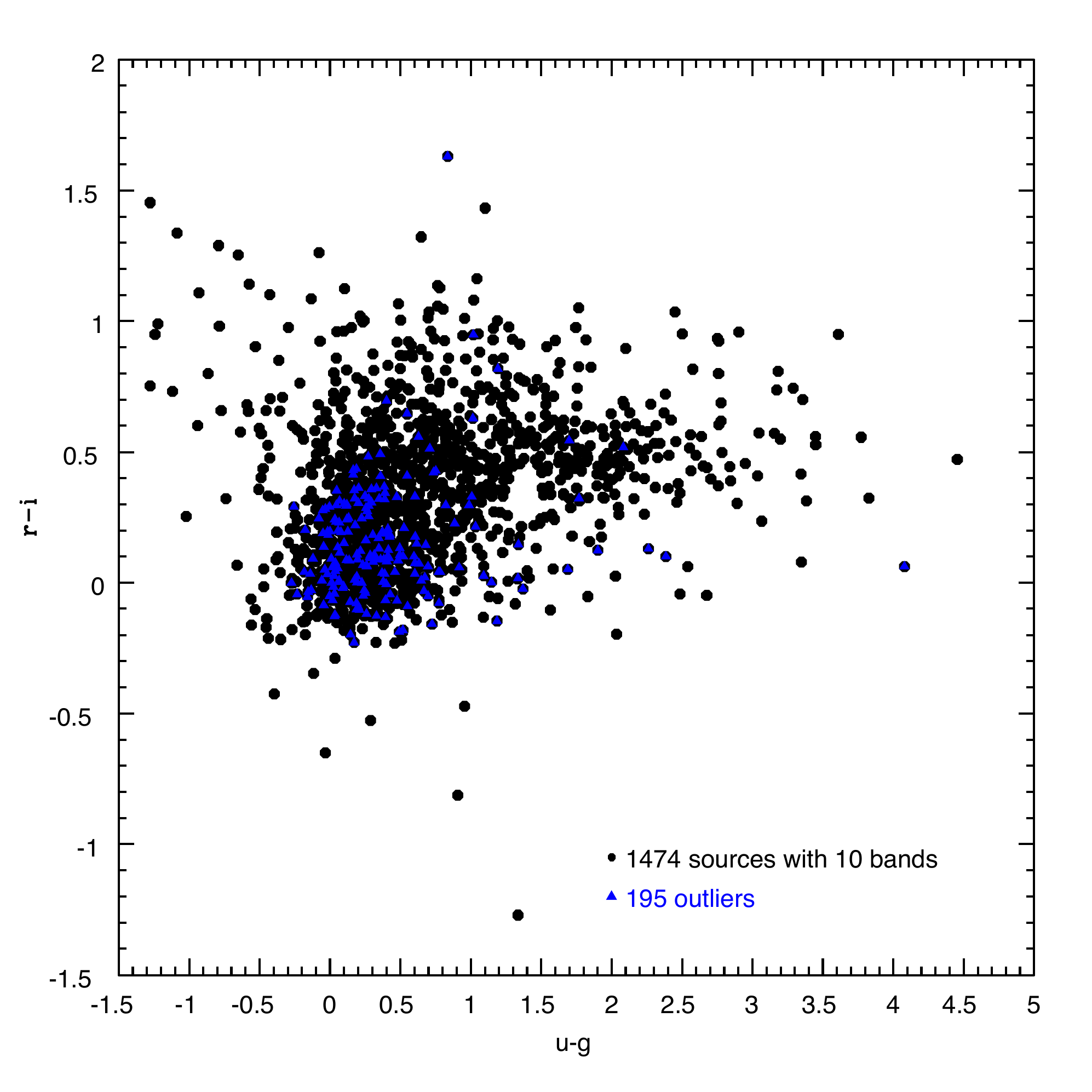}
\includegraphics[height=1.\columnwidth]{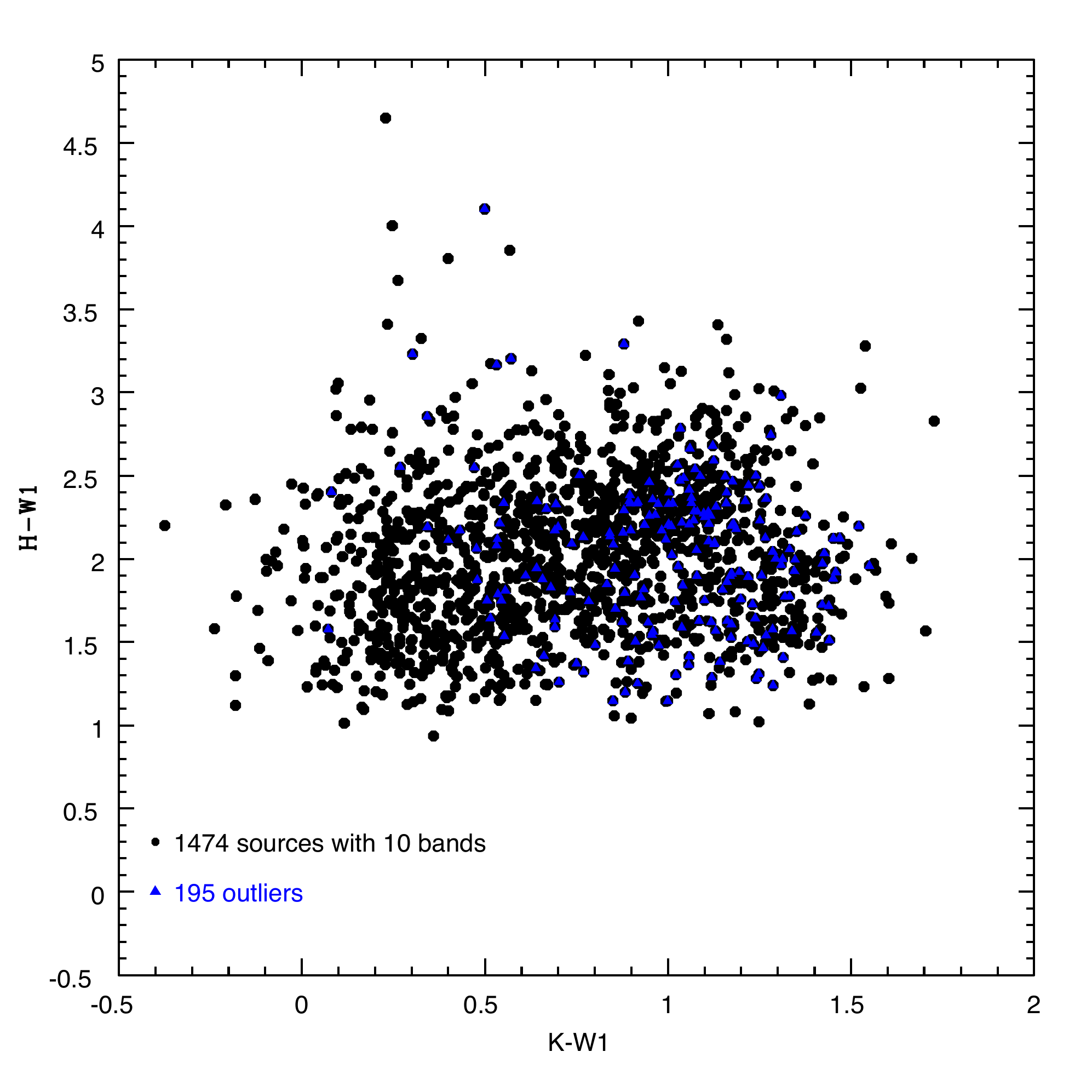}
\caption{{\bf  Left:} The r-i vs. u-g colour distribution of the training sample (black circles) and the outliers (blue triangles). {\bf Right:} The H-W1 vs. K-W1 colour distribution of the training sample (black circles) and the outliers (blue triangles).}
\label{fig_colours_outliers}
\end{figure*}

\begin{figure}
\centering
\includegraphics[width=\hsize]{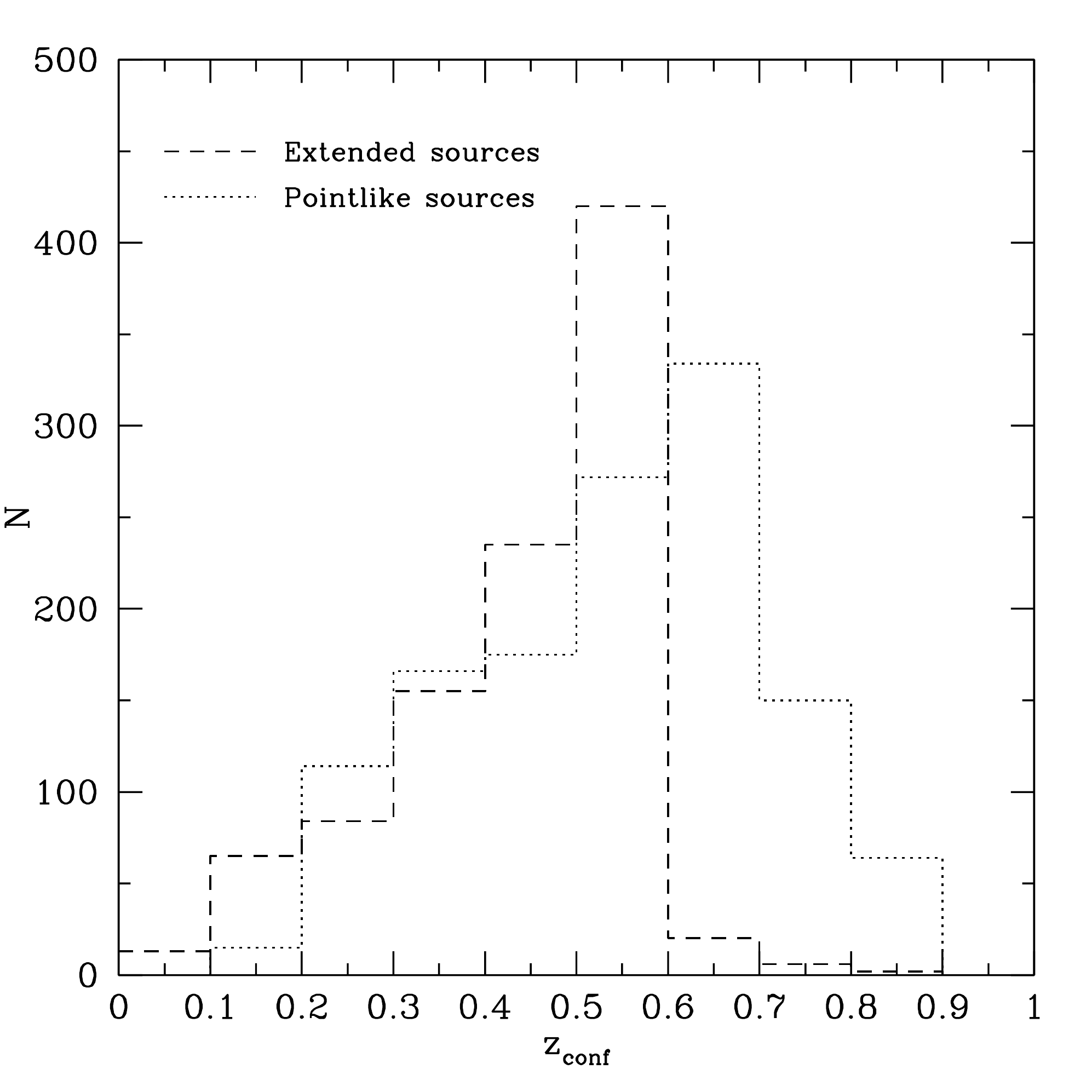}
\caption{The distribution of $z_{conf}$ for the extended (dashed line) and the pointlike (dotted line) sources in our training sample.}
\label{fig_zconf}
\end{figure}

\section{Analysis}

\subsection{Methodology}
For the estimation of the photometric redshifts for the X-ray AGN in the ATLAS field we used the publicly available algorithm named TPZ (Trees for Photo-z). The technique is described in detail in Kind \& Brunner 2013. In brief, TPZ is a parallel, machine learning algorithm that uses prediction trees and random forest techniques to generate photometric redshift Probability Density Functions (PDFs), by incorporating into the calculation measurement errors while also dealing efficiently with missing values in the data. 

Random forest is an ensemble learning method for classification, regression and other tasks that generates prediction trees and then combines their predictions together. Prediction trees are built by asking questions that split the data until a stopping criterion is met and that creates a terminal leaf. The leaf contains a subsample of the data with similar properties and by applying a model within the leaf a prediction is made. 

TPZ is an empirical technique and therefore requires a dataset with spectroscopically measured redshifts to train the algorithm before it is applied to our photometric X-ray sample. The spectroscopic, training sample we used in our analysis is described in the following Section. 

\subsection{Training Sample}

The X-ray catalogue we use to train the TPZ algorithm comes from the XXM-XXL survey. XMM-XXL covers a total of about 50\,deg$^2$ with an exposure time of about 10\,ks per XMM pointing \citep{Liu2016}. 8,445 X-ray sources are detected in the north region (XXL-N) that extends to about 25\,deg$^2$. 5,294 of these sources have optical (SDSS) photometry. Reliable spectroscopy from SDSS-III/BOSS is available for 2,512 AGN \citep{Menzel2016}. To increase the size of our training sample we also include sources from the XWAS \citep[XMM-Newton Wide Angle Survey;][]{Esquej2013}, XBS \citep[][]{DellaCeca2004}, XMS \citep[][]{Barcons2007} and COSMOS \citep[][]{Brusa2010} surveys. We also add $\sim 1,500$ optically selected X-ray AGN with spectroscopic redshifts from the SDSS-DR13 dataset, by cross-matching the 3XMM-DR5 catalogue with SDSS, UKIDSS \citep[][]{Hambly2008, Irwin2009}, 2MASS \citep[][]{Skrutskie2006} and WISE. This increases the total number of sources in our training sample to 5,157 (Table \ref{table_training_numbers}). Testing the performance of TPZ (see next Section) with and without the optically selected X-ray AGN revealed that the inclusion of these extra sources marginally but systematically improves the training process of the TPZ code. Specifically, the outliers percentage (see next Section) goes down by 2-3\% in all cases. Therefore, the results we present next are estimated using the training sample described above.

Additionally to the photometric bands of SDSS (u, g, r, i, z) we also include mid-IR (W1, W2)  and near-IR (J, H, K) bands in the training process of TPZ to check whether its performance improves. For that purpose we cross-match the 5,157 sources with the WISE catalogue and near-IR catalogues, i.e. VISTA, UKIDSS or 2MASS. The cross-match was performed using the xmatch cross-correlation tool and following the same analysis described in the previous Section for the ATLAS sources. The number of sources we obtained and the available photometry is presented in Table \ref{table_training_numbers}. Although TPZ can infer missing photometry, in our validation tests and the estimation of the photometric redshifts of the X-ATLAS sources, only the available photometric bands are used for each subsample. 

The redshift distribution of the training set is presented in Fig. \ref{fig_n_z}.

\begin{figure}
\begin{center}
\includegraphics[height=0.77\columnwidth]{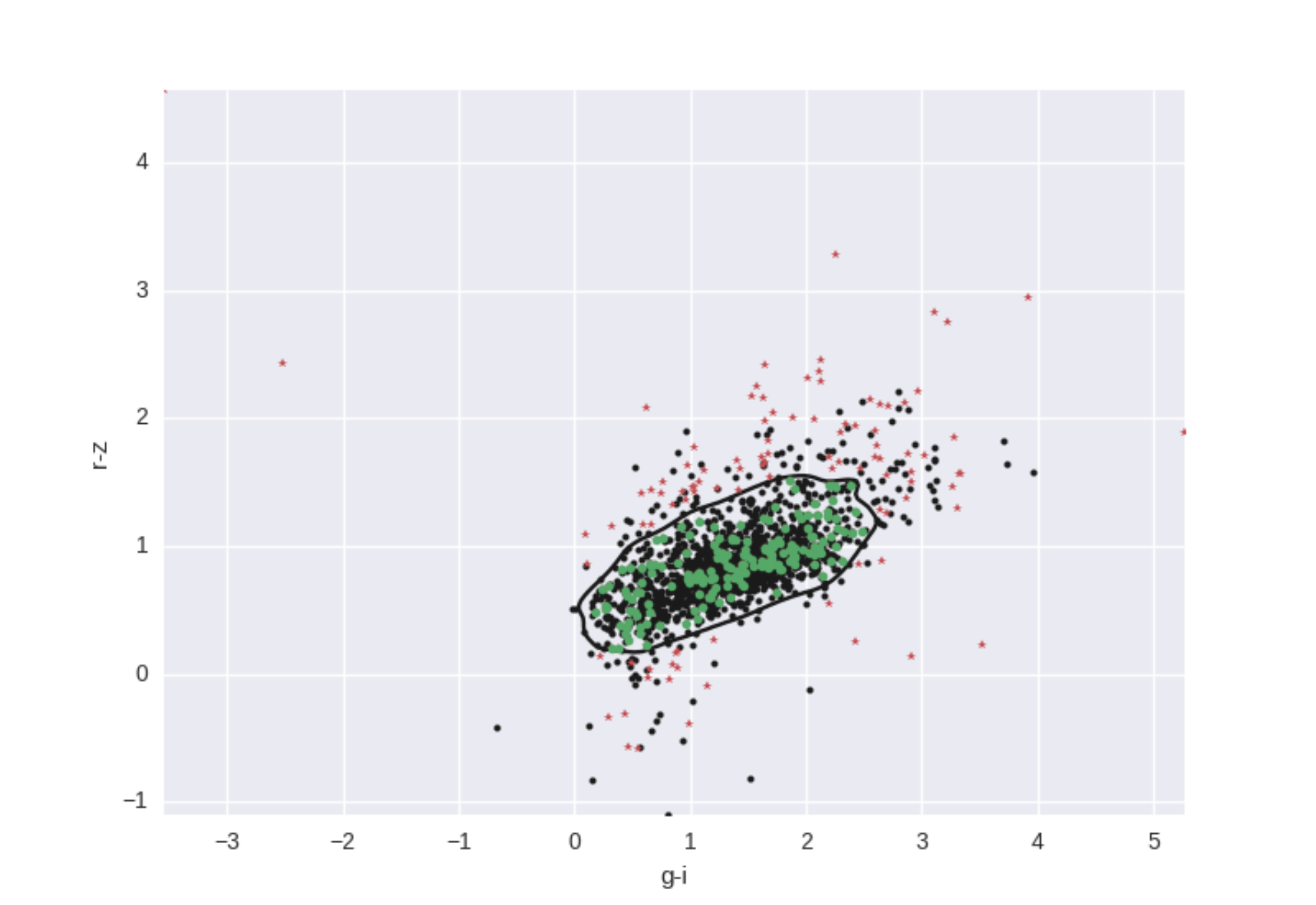}
\end{center}
\caption{i-z vs. g-i colour space diagram. Black dots present the sources in our training sample. The black solid line defines the region of the colour space that contains  90\% of the training sources as estimated  by the KDE test. Green dots are the sources from the X-ATLAS sample inside the 90\% region and red crosses present the remaining of the X-ATLAS sources. }
\label{fig_colour_kde}
\end{figure}

\begin{figure}
\begin{center}
\includegraphics[height=1.\columnwidth]{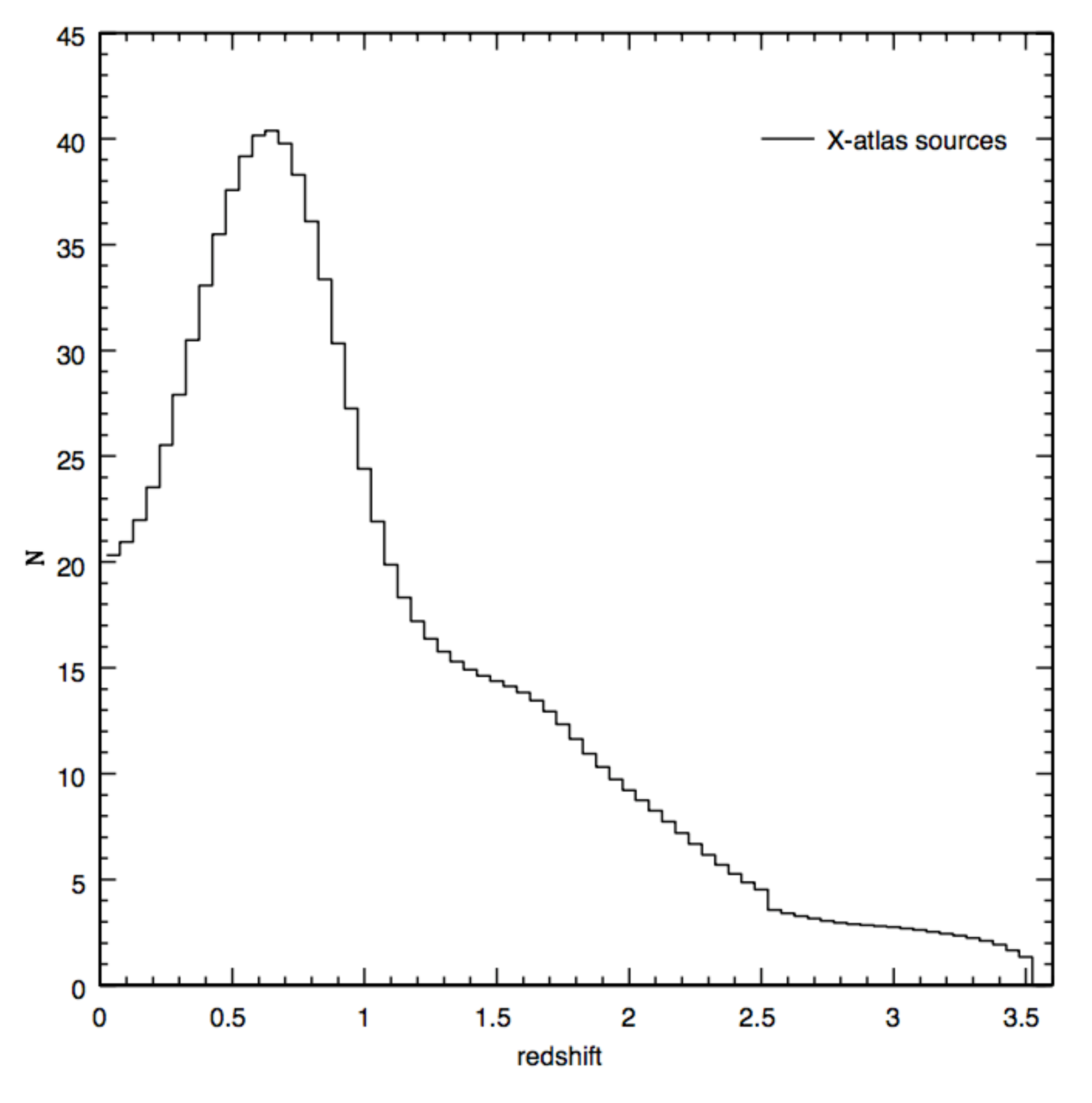}
\end{center}
\caption{The redshift distribution of the 933 X-ATLAS sources taking into account the full PDF of each source. Photoz are estimated using the TPZ algorithm.}
\label{fig_n_z_atlas}
\end{figure}

\begin{figure*}
\begin{center}
\includegraphics[height=1.\columnwidth]{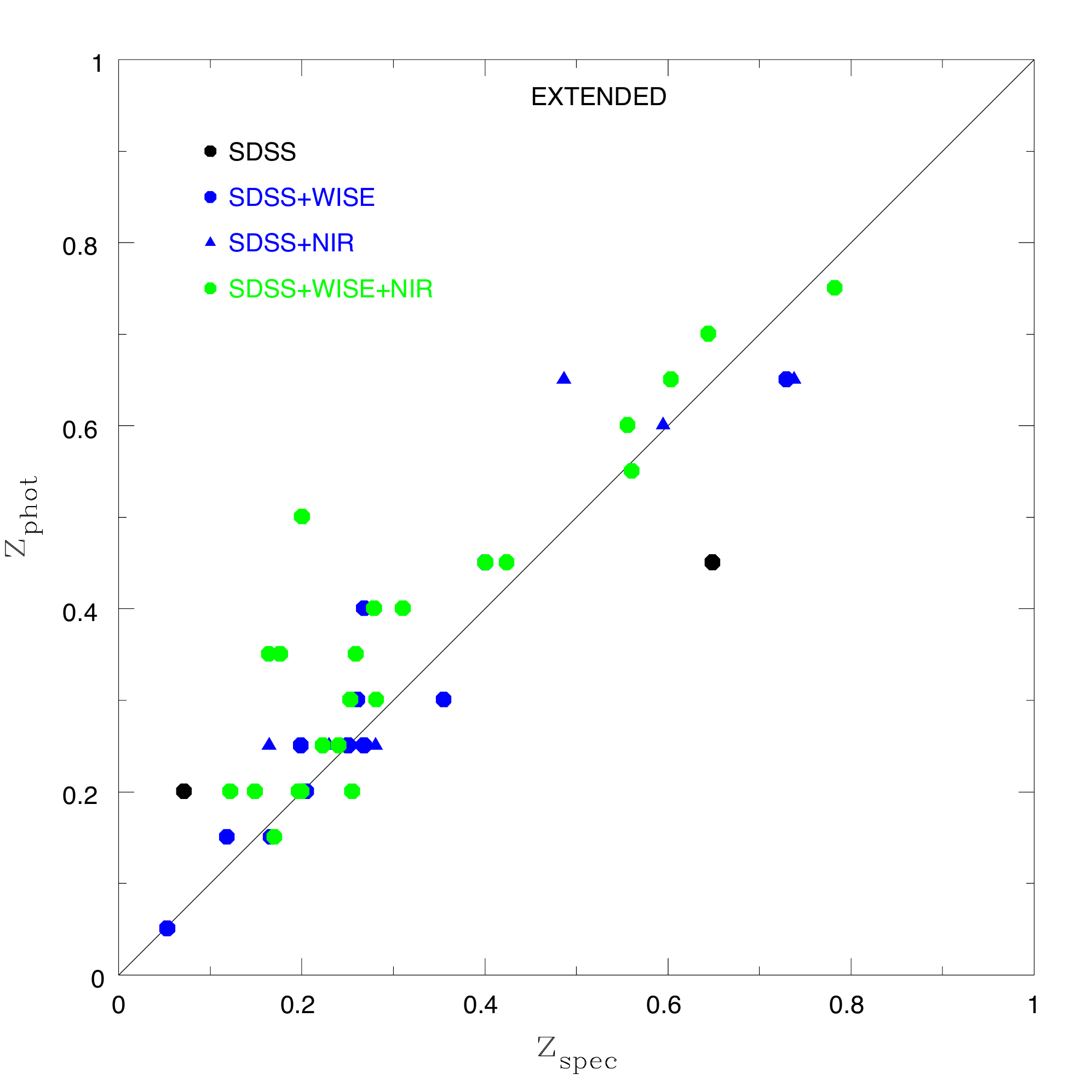}
\includegraphics[height=1.\columnwidth]{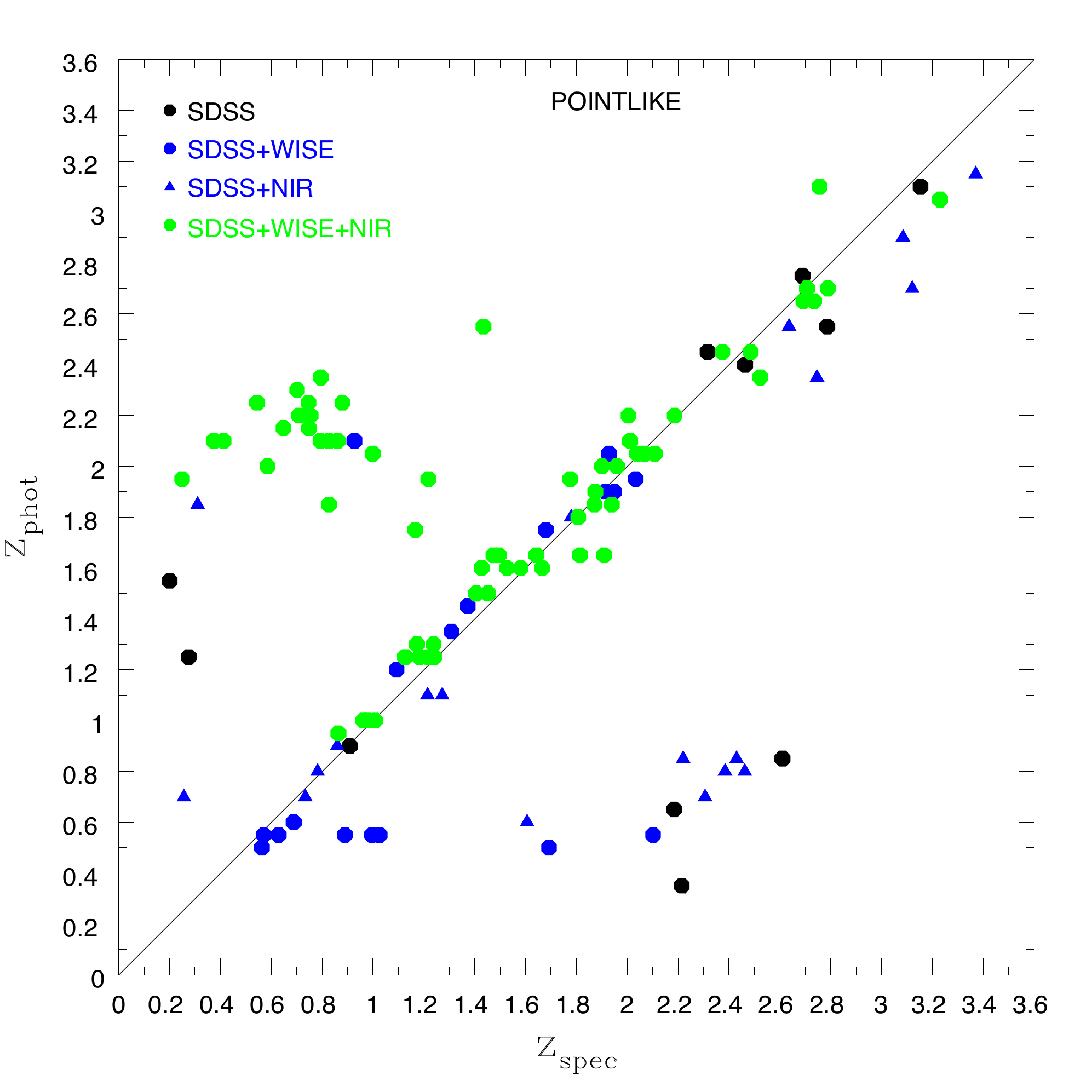}
\end{center}
\caption{Comparison of the photometric redshifts estimated using TPZ with the spectroscopic redshifts from the SDSS and GAMA surveys for the 174 out of the 933 sources in the ATLAS field. The left panel shows the comparison for 55 extended sources and the right panel for 119 point-like sources. The median error of the photoz varies from 0.19 to 0.26 and the median confidence level from 0.36 to 0.49, depending on the morphology of the source and the available photometric bands (Table \ref{table_atlas_errors}). A significant fraction of outliers exist in the case of pointlike source, even when seven or even 10 photometric bands are used. This number can be greatly reduced if a cut is applied on the confidence level of the photometric redshift, as discussed in the text ($z_{conf}>0.6$).}
\label{fig_zcomp_tpz}
\end{figure*}

\begin{figure}
\begin{center}
\includegraphics[height=1.\columnwidth]{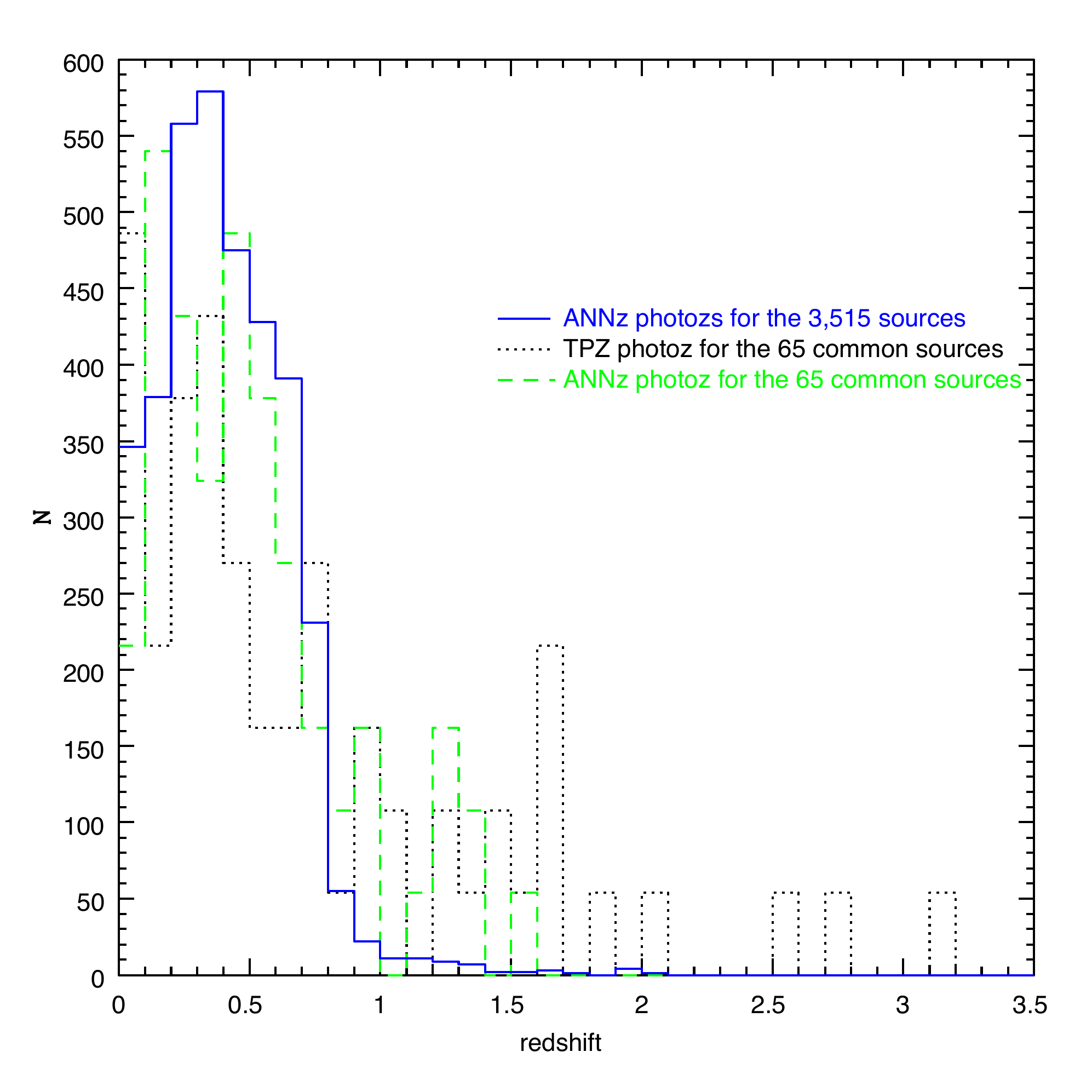}
\end{center}
\caption{The redshift distribution of the 3,515 sources with ANNz photoz estimation in the X-ATLAS field (solid line) and the N(z) using TPZ (normalized to the number of sources with ANNz estimation), of the 65 sources that also belong to our X-ray AGN sample. The redshift distribution of the photoz estimated by ANNz peaks at low redshifts ($z\sim 0.3$) and there is a very small number of sources with $z>1$ (solid line). This is expected since ANNz has been trained to estimate photometric redshifts for galaxies. The N(z) estimated using TPZ has been specifically trained to estimate photoz for X-ray sources and presents a second peak at $z\sim 1.5$ (dotted line).}
\label{fig_n_z_annz}
\end{figure}

\begin{figure}
\begin{center}
\includegraphics[height=1.\columnwidth]{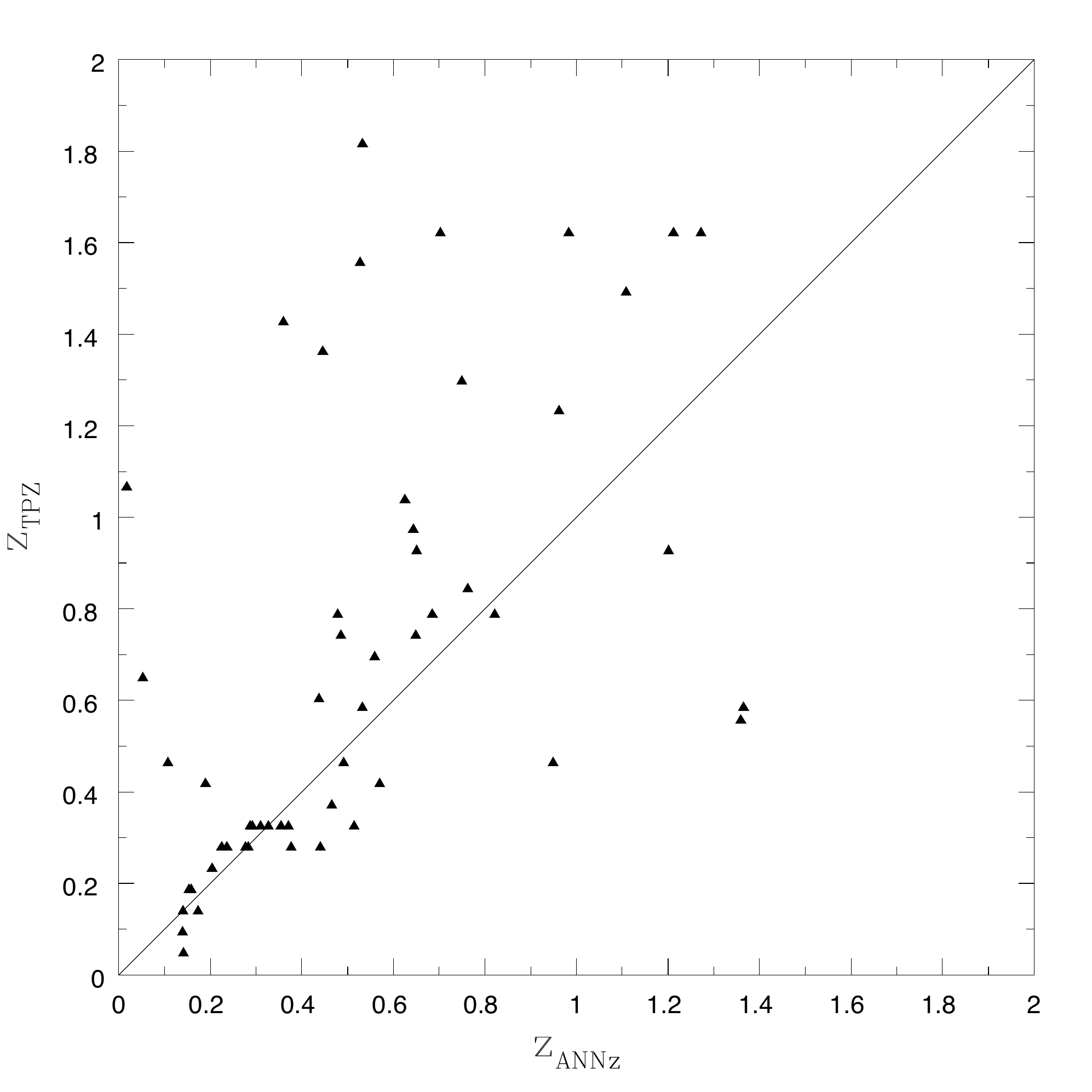}
\end{center}
\caption{Comparison of our photometric redshifts estimated using TPZ and those estimated using ANNz (Smith et al. 2011) for the 65 common sources with our X-ATLAS X-ray AGN catalogue and the submillimetre catalogue described in Valiante et al. 2016 and Bourne et al. 2016. Most of the discrepancy between the two photoz estimations is located in the upper left part of the plot, i.e., ANNz computes lower redshift values compared to our TPZ measurements. Most of this difference is likely due to the different training sets used in the two methods. The training sample of ANNz is constructed to better suit their test sample, the vast majority of which consists of galaxies. Our training sample (Section 3.2) consists of X-ray AGN (see  text for more details).}
\label{fig_zcomp_annz}
\end{figure}

\subsection{Checking the performance of TPZ using the training set}

To check the performance of TPZ in estimating accurate photometric redshifts, we split our training set into two subsamples. One is used to train the algorithm and the other subsample is used as a test case for which we estimate photometric sources. This is an ideal scenario since both subsamples share the same region of the parameter space and the same quality of (spectroscopic) data, i.e. the same distribution in redshift and magnitude as well as the same photometric errors. To account for the fainter magnitudes of our photometric X-ATLAS sources compared to the spectroscopic, training sample and facilitate a more accurate check of the TPZ performance, in this test we shall train TPZ using colours instead of magnitudes. Fig. \ref{fig_colours} presents two examples of the colour distribution of the training sources (black circles).

The accuracy of the photometric redshifts estimated by TPZ is quantified by two widely used statistical parameters, the normalized absolute median deviation, $\sigma_{\rm{nmad}}$, and the percentage of outliers, $\eta$.  $\sigma_{\rm{nmad}}$ is defined as:

\begin{equation*}
\Delta (\rm{z_{norm}})=\frac{z_{spec}-z_{phot}}{1+z_{spec}},
\end{equation*}
\begin{equation*}
\begin{aligned}
MAD(\Delta (\rm{z_{norm}}))=\rm{Median}(|\Delta (\rm{z_{norm}})| ),
\end{aligned}
\end{equation*}
\begin{equation}
\sigma_{\rm{nmad}}=1.4826 \times MAD(\Delta (\rm{z_{norm}})). 
\end{equation}

The percentage of outliers, $\eta$, is defined as:

\begin{equation}
\eta=\frac{100}{N} \times (\rm{Number\, of\, sources\, with\,} |\Delta (z_{norm})| > 0.15)
\end{equation}

Since the near-IR data come from different surveys, the training sample is used to calibrate any possible dependencies on the different filters used, i.e. differences between the K filter on UKIDSS and the K$_s$ filter on VISTA and 2MASS. Our tests reveal that there are no differences whether we ignore the different filters or we scale K magnitudes to K$_s$. For example using the SDSS+NIR sample, for pointlike and extended sources, the percentage of outliers differs by $<\pm0.8\%$ and the difference in $\sigma_{nmad}$ is negligible. Therefore, we ignore this difference in filters in our analysis. 

Our initial tests reveal that the performance of the TPZ algorithm in estimating photometric redshifts improves when we split the sources based on their morphology. Using the SDSS photometric bands and estimating photometric redshifts without dividing the sources into pointlike and extended we get, $\sigma_{nmad}=0.12$ and $\eta=0.35\%$. These numbers are higher than those derived when splitting the sources based on their optical morphology (see Table \ref{table_performance}). We also try to use the morphology as one of the features used to train the algorithm. Our tests reveal that there is no improvement in the accuracy of the photoz estimations. For example, using 10 photometric bands, $\sigma_{\rm{nmad}}=0.05$ and $\eta= 11.8\%$. These estimations are in-between the values obtained when the sources are split based on their morphology (Table \ref{table_performance}). We therefore split the training sources into point-like and extended, using their SDSS classification. The number of sources in each subsample is shown in Table \ref{table_training_numbers}. Their redshift distribution is presented in Fig. \ref{fig_n_z}. Based on the two distributions we can reach redshifts up to 3.5 and 2.5 for point-like and extended sources, respectively.

{Table \ref{table_performance} presents the values for the various parameters of TPZ used to estimate photometric redshifts for each subsample. Nrandom is the number of random realizations that TPZ performs, NTrees is the number of trees used and Natt the number of attributes for TPZ. The number of the bins used is 50 in the case of extended sources and 70 for the pointlike sources. For the estimation of the PDFs and the confidence level of the estimated photometric redshifts \citep[see][]{Kind2013} the rms factor is set to 0.06. The same values for each parameter are used for the estimation of the photoz for the 1,031 X-ray sources in the ATLAS field (next Section). 

Fig. \ref{fig_importance_point} presents the importance of some of the attributes used in the training process of the TPZ algorithm. The  left panel presents the importance of the attribute  as a function of redshift for the pointlike sources when ten photometric bands are available. A factor of one in the importance implies that the attribute acts as a random variable \citep[for more details see Section 4.1.1. in][]{Kind2013}. The right panel presents the RMS importance factor as a function of the attributes computed by using the bias, defined as $\Delta z = z_{spec}-z_{phot}$, and its scatter. Fig. \ref{fig_importance_extended} shows the same measurements for the extended sources.

The left panels of Figures \ref{fig_importance_point} and \ref{fig_importance_extended} show that the importance of each attribute is different at different redshifts. In the case of pointlike sources, the z$-$W1 colour is the most important attribute up to redshift 2.5, but its importance significantly drops at $z=3$. Similarly, the importance of the h-k colour in the case of extended sources significantly drops at $z>1.4$. Moreover, same colours have different importance for pointlike and extended sources, as can be more clearly seen in the right panels of the two Figures. For instance, the z-W1 colour is the most important attribute for the pointlike sources but the least important one in the case of extended sources. Therefore, the importance of the colours used to estimate photometric redshifts for X-ray sources strongly depends on the morphology of the source and the redshift range of interest.

The results of our measurements are presented in Table \ref{table_performance}. Using only optical photometry (SDSS) the number of outliers is high, especially in the case of point-like sources. Adding mid-IR colours (WISE) the results improve significantly while TPZ performs best when we also include near-IR magnitudes in the training process of the algorithm. Fig. \ref{fig_test_tpz} compares the estimated photometric redshifts with the available spectroscopic redshifts of the sources. Fig. \ref{fig_pdf_examples} presents examples of photometric redshift PDFs produced by TPZ.

The number of outliers drops down to 9-14\%  when 10 bands are used for the photoz estimation (Table \ref{table_performance}). Although this number is significantly lower compared to the outliers percentage we get using fewer number of photometric bands, there is a non-negligible number of outliers even among our best photoz estimations.  Fig. \ref{fig_colours_outliers} presents the colour space occupied by the training sample (black circles) for different colour combinations. Outliers (blue triangles) lie within the boundaries of the training set. Therefore, their existence cannot be attributed to extrapolation in colour space that TPZ may be  required to perform. Although the cause of these outliers is uncertain, their percentage can be significantly reduced if a cut is applied in the confidence level, $z_{conf}$ \citep[][]{Kind2013}, of the photoz. For example, for $z_{conf}>0.6$, $\eta=4.5\%$ in the case of pointlike sources. The percentage goes further down ($\eta=2.4\%$) when we consider only photoz estimated using 10 photometric bands. Applying a $z_{conf}>0.5$ cut for the extended sources the corresponding numbers are, $\eta=4.0\%$ and  $\eta=1.2\%$. Fig. \ref{fig_zconf} presents the distribution of $z_{conf}$ for pointlike and extended sources.

Variability of AGN can impact the accuracy of the estimated photometric redshifts \citep{Simm2015}. This is not an issue for the optical bands of SDSS we use, since all bands have been observed simultaneously. Variability is also minimal in the mid-IR photometric bands. Regarding the near-IR bands, though, an estimation of the variable sources in our sample cannot be made. We would expect most of these sources to be excluded when a $z_{conf}$ cut is applied, as discussed above, but a flag cannot be assigned to indicate these sources in the full catalogue.

\begin{table}
\caption{The number of sources used to train TPZ, with the corresponding available photometry. The second column presents the total number of the sources, while the third and fourth columns show the numbers of sources divided into point-like and extended. In the parentheses we quote the number of the sources used to train TPZ during the validation process (see text for more details).}
\centering
\setlength{\tabcolsep}{0.7mm}
\begin{tabular}{cccc}
       \hline
 \hline
Available photometry & Total number of & Point-like & Extended \\
& sources &sources&sources \\
       \hline
SDSS & 5157 & 2703 (1900) & 2454 (1200)    \\
SDSS+WISE  &4781 & 2473 (1500) & 2308 (1400)   \\
SDSS+WISE+NIR & 3212 & 1613 (1000) & 1599 (1000)  \\
SDSS+NIR& 3313 & 1679 (1000) & 1634 (1100)   \\
\hline
\label{table_training_numbers}
\end{tabular}
\end{table}

\begin{table*}
\caption{The performance of the TPZ algorithm, estimated by splitting our spectroscopic sample (see Section 3.2) into train and test files. The accuracy of the photometric redshifts is quantified by the estimation of the normalized absolute median deviation, $\sigma_{\rm{nmad}}$ and the percentage of outliers, $\eta$. The median error of the photometric redshift for each subsample is shown. The values of the TPZ parameters used for each subsample is also presented.}
\centering
\setlength{\tabcolsep}{2mm}
\begin{tabular}{cccccccc}
       \hline
 \hline
Sample &  \multicolumn{2}{c}{Point-like} & \multicolumn{2}{c}{Extended} & \multicolumn{3}{c}{TPZ parameters} \\
& $\sigma_{\rm{nmad}}$ / $\eta$ (\%)  & <error> &   $\sigma_{\rm{nmad}}$ / $\eta$ (\%)  & <error> &Nrandom & NTrees & Natt\\
       \hline
\\       
SDSS & 0.08 / 27.0 &  0.33 & 0.06 / 18.0 &  0.21 &  6 & 8 & 7\\
SDSS+WISE & 0.06 / 17.4 & 0.25 & 0.06 / 13.0 & 0.20 & 8 & 10 & 8 \\
SDSS+WISE+NIR &  0.05 / 13.7 & 0.23  &0.04 / 9.0 & 0.18 & 6 & 8 & 12\\
SDSS+NIR & 0.06 / 20.0 & 0.27 & 0.05 / 11.5 & 0.19 & 8 & 10 & 10 \\
\hline
\label{table_performance}
\end{tabular}
\end{table*}

\section{Results}

Following the results of the tests during the validation process (see previous Section), we split the 1,031 X-ATLAS X-ray AGN into point-like and extended sources, using their SDSS classification. The number of sources divided based on their optical morphology as well as the available photometry is presented in Table \ref{table_agn_numbers}. 

Machine learning methods, such as TPZ, are known to perform poorly when training set coverage is not available and extrapolation must be performed \citep[][]{Beck2017}. Fig. \ref{fig_colours} compares the colour distribution of the X-ATLAS AGN (blue triangles) with that of the training sample (black circles). In both examples, the coverage of the training set seems sufficient to properly train TPZ for the estimation of the photometric redshift of the X-ATLAS sources. To quantify the differences among the colours between the training and the X-ATLAS samples we perform a Kernel Density Estimation (KDE) test. Using KDE we define the region in colour space that contains 90\% of the training sample. Then we estimate the fraction of the X-ATLAS sources that are contained in that region, i.e. these sources are well covered by the training sample. This is illustrated in Fig. \ref{fig_colour_kde} for the $g-i$ vs $r-z$ colours.
Table \ref{table_colours_kde}  presents the fraction of X-ATLAS sample that is well covered in all possible combinations of colours as well as in at least one colour-colour combination.

TPZ estimates photoz for 933 out of the 1,031 sources. Most of the remaining 98 sources have missing photometry, i.e., only SDSS bands are available and therefore the algorithm cannot be properly trained to give a photometric redshift estimation. The distribution of the photometric redshifts for the 933 X-ATLAS X-ray sources, estimated by TPZ and taking into account the full PDF of each sources, is shown in Fig. \ref{fig_n_z_atlas}. Out of the 933 AGN, 174 have available spectroscopic redshifts from the SDSS and GAMA surveys. In Fig. \ref{fig_zcomp_tpz} we compare our photometric redshifts, estimated using TPZ with the available spectroscopic redshifts. Table \ref{table_atlas_errors} presents the median error and the median confidence level, $z_{conf}$, of the photometric redshifts, calculated by TPZ as a function of the available photometric bands. The full catalogue with the estimated photometric redshifts is available online\footnote{\textrm{http://xraygroup.astro.noa.gr/atlas/atlas-photoz-online.dat}}.

To check how many of the X-ATLAS sources are AGN ($\rm{log}\,L_X > 42$\,erg\,s$^{-1}$) we use the X-ray fluxes provided by the XMM-ATLAS catalogue \citep[][]{Ranalli2015} and the estimated photometric redshifts to calculate the X-ray luminosities. This information is available for 894 sources. Our calculations show that 883 of the sources have $\rm{log}\,L_X > 42$\,erg\,s$^{-1}$.

\begin{table*}
\caption{The fraction of X-ATLAS sample that is well covered in all possible combinations of colours as well as in at least one colour-colour combination. An X-ATLAS source is considered well covered by the training set, in a colour-colour combination, when it lies in a region of the colour space that contains 90\% of the training sources.}
\centering
\setlength{\tabcolsep}{0.7mm}
\begin{tabular}{ccc}
       \hline
 \hline
Available photometry & Fraction of sources well-covered &  Fraction of sources  well-covered\\
&  in all colour combinations & in at least one colour-colour combination \\
\hline
& Extended/Pointlike & Extended/Pointlike \\
       \hline
SDSS & 51\% / 56\% & 98\% / 91\%   \\
SDSS+WISE  &40\% / 44\% & 99\% / 94\%   \\
SDSS+WISE+NIR & 25\% / 35\% & 100\% / 100\%  \\
SDSS+NIR& 37\% / 46\% & 99\% / 99\% \\
\hline
\label{table_colours_kde}
\end{tabular}
\end{table*}

\begin{table}
\caption{The median error of the photometric redshifts and their median confidence level, estimated by TPZ, for each subsample of the X-ATLAS dataset, based on the available photometry}
\centering
\setlength{\tabcolsep}{0.7mm}
\begin{tabular}{ccc}
       \hline
 \hline
Available photometry & <$z_{conf}$> & <error> \\
\\
& Extended/Pointlike & Extended/Pointlike \\
       \hline
SDSS & 0.44 / 0.36 & 0.21 / 0.26   \\
SDSS+WISE  & 0.44 / 0.46 & 0.20 / 0.25   \\
SDSS+WISE+NIR & 0.49 / 0.48 & 0.19 / 0.24  \\
SDSS+NIR& 0.48 / 0.47 & 0.19 / 0.26 \\
\hline
\label{table_atlas_errors}
\end{tabular}
\end{table}

\section{Summary and Discussion}

In this paper, we present a catalogue with photometric redshift estimations for 933 X-ray AGN in the ATLAS field. For the first time, we have used the largest available X-ray sample to train a machine learning technique (TPZ) and estimate photo-z for X-ray sources. Our analysis shows that our redshift estimations are accurate when optical photometry is combined with mid-IR photometry in the training process of the algorithm. Using additional photometric bands (near-IR) further improves the precision of photometric redshifts.  Our photo-z estimations have a normalized absolute median deviation, $\sigma _{\rm{nmad}}\approx$0.06 and the percentage of outliers is, $\eta$=10-14\%, depending on whether the sources are extended or point-like. These numbers significantly improve when a cut in the confidence level of the photometric redshift is applied ($z_{conf}>0.5-0.6$.).

\cite{Valiante2016} and \cite{Bourne2016} presented a catalogue of 120,230 sources with identification of optical counterparts to submillimetre sources in Data Release 1 (DR1) of the H-ATLAS sample. The sources are located in three fields on the celestial equator, covering a total area of 161.6\,deg$^2$, previously observed in the GAMA spectroscopic survey. The catalogue contains photometric redshifts \citep[][]{Smith2011} measured from the SDSS {\it{ugriz}} and UKIDSS {\it {YJHK}} photometry using the neural network technique of ANNz \citep{Collister2004}. Photometric redshifts have been estimated using a training sample constructed by spectroscopic redshifts from GAMA I, SDSS DR7, 2SLAQ \citep[][]{Cannon2006}, AEGIS \citep[][]{Davis2007} and zCOSMOS \citep[][]{Lilly2009} covering redshifts $z<1$. 5500 of these sources lie in the X-ATLAS region and 3,515 have a photometric redshift estimation using ANNz. 65 of these sources are common between the two samples. Fig. \ref{fig_n_z_annz} presents the redshift distribution of the 3,515 sources (solid line) and that of the 65 common sources, based on our TPZ photoz estimations (dashed line). The vast majority of the ANNz photoz estimations are at $z<1$ due to the galaxy training sample used for ANNz. In Fig. \ref{fig_zcomp_annz} we compare our photometric redshift estimations using TPZ with those using the ANNz method. Most of the discrepancy between the two photoz estimations is located in the upper left part of the plot, i.e., ANNz computes lower redshift values compared to our TPZ measurements. Most of this difference is likely due to the different training sets used in the two methods. The training sample of ANNz is constructed to better suit their test sample, the vast majority of which consists of galaxies. Our training sample (Section 3.2) consists of X-ray AGN and extends to higher redshifts (up to z$\sim$3.5; see Fig. \ref{fig_n_z}). Our analysis has shown (Figures \ref{fig_importance_point}, \ref{fig_importance_extended}, \ref{fig_colour_kde} and Table \ref{table_colours_kde}), that the coverage of our training set in feature space , i.e., colours, is also sufficient at high redshifts ($z>1$). The results of this comparison is not an indication that ANNz generally performs poorer compared to TPZ, but that for the specific X-ray sources our X-ray training set is probably better suited.

Large-scale structure studies (e.g. weak lensing, gravitational waves, clustering) require accurate redshifts in their analysis. \cite{Georgakakis2014} examined how the accuracy of photometric redshifts affects the estimation of the correlation function in clustering measurements. They concluded that a $\sigma \sim 0.04$ (standard deviation of the photoz) is required in photo-z estimations to be used for the calculation of the AGN correlation function in clustering studies. This accuracy is challenging to obtain, albeit, Georgakakis et al. argue that the clustering signal can be recovered even if the normalized absolute median deviation is $\sigma=0.08$, when the AGN/galaxy cross-correlation function is measured and the galaxy sample has very accurate photometric redshifts ($\sigma \approx 0.01$). Their analysis takes into consideration the error of the photometric redshifts but does not account for outliers. Even our best photometric redshift measurements (extended sources with ten photometric bands available) have a considerable percentage of outliers ($\sim$ 9-10$\%$). Our preliminary results (Mountrichas et al., in prep.) indicate that the clustering signal can be recovered using photometric redshifts derived by TPZ, when a cut is applied on the confidence level of the photometric redshift.

The 3XMM catalogue is the largest X-ray catalogue available, containing about 470,000 unique sources covering a total area of 1,000\,deg$^2$ on the sky. XMMFITCAT-Z \citep[http://xraygroup.astro.noa.gr/Webpage-prodec/xmmfitcatz.html;][]{Corral2015} is a spectral fit database for 124,000 sources with good photon statistics in the 3XMM. The potential of these catalogues will increase significantly with the addition of the distance information for their sources. We shall apply the analysis presented in this work on the 3XMM catalogue, to estimate photometric redshifts for all the X-ray sources with, at least, optical photometry available. In the 3XMM-DR5 catalogue, 42,697 sources have available SDSS photometry and 22,619 have also WISE counterparts. 3XMM-DR6 and usage of PanSTARRS in the southern sky will increase the numbers of available X-ray sources. The resulting X-ray catalogue will exceed, by an order of magnitude, any other X-ray catalogue with available redshift information, up to date.

\begin{acknowledgements}
The authors thank the anonymous referee for their careful reading of the paper and their constructive comments. The research leading to these results has received funding from the European Union's Horizon 2020 Programme under the AHEAD project (grant agreement n. 654215). GM acknowledges financial support from the AHEAD project that is funded by the European Union as Research and Innovation Action under Grant No: 654215. FJC and ACR acknowledge financial support through grant AYA2015-64346-C2-1-P (MINECO/FEDER). ACR also acknowledges financial support
by the European Space Agency (ESA) under the PRODEX program.
\end{acknowledgements}

\bibliography{mybib}{}
\bibliographystyle{aa}

\end{document}